\documentclass[11pt,nofootinbib]{article}
\usepackage{jheppub}
\usepackage[utf8x]{inputenc}
\pdfoutput=1
\usepackage{graphicx}
\usepackage{subfigure}
\usepackage{amsmath}
\usepackage{bm} 
\usepackage{xspace}
\usepackage{relsize}
\usepackage{placeins}
\usepackage{multirow}
\usepackage{soul}

\def\babar{\mbox{\slshape B\kern-0.1em{\smaller A}\kern-0.1em 
  B\kern-0.1em{\smaller A\kern-0.2em R}}}
\usepackage{lmodern}

\usepackage{color}
\usepackage[usenames,dvipsnames,svgnames,table]{xcolor}
\usepackage{hyperref} 
\hypersetup{linktocpage=true}
\usepackage[all]{hypcap}

\numberwithin{equation}{section}
\newcommand{\beq}{\begin{equation}}
\newcommand{\eeq}{\end{equation}}
\newcommand{\bea}{\begin{eqnarray}}
\newcommand{\eea}{\end{eqnarray}}
\newcommand{\bear}{\begin{eqnarray}}
\newcommand{\eear}{\end{eqnarray}}

\newcommand{\ba}{\begin{array}}
\newcommand{\ea}{\end{array}}

\def\gev{\,{\rm GeV}}

\newcommand{\met}{\slash{\hspace{-2.5mm}E}_T}
\newcommand{\metp}{\slash{\hspace{-2.5mm}p}_T}
\def\fbi{\,{\rm fb}^{-1}}
\def\abi{\,{\rm ab}^{-1}}

\begin{document}

\title{\boldmath 
  Heavy Higgs as a Portal to the Supersymmetric Electroweak Sector}
  \vspace*{0.5cm}

\author[a,b]{Stefania Gori,}
\author[c,d]{Zhen Liu,}
\author[a,b,e,f]{Bibhushan Shakya}

\affiliation[a]{Santa Cruz Institute for Particle Physics, University of California, Santa Cruz, CA 95064, USA}
\affiliation[b]{Department of Physics, 1156 High St., University of California Santa Cruz, Santa Cruz, CA 95064, USA}
\affiliation[c]{Maryland Center for Fundamental Physics, Department of Physics, University of Maryland, College Park, MD 20742, USA}
\affiliation[d]{Theoretical Physics Department, Fermilab, Batavia, IL 60510, USA}
\affiliation[e]{Department of Physics, University of Cincinnati, Cincinnati, Ohio 45221, USA}
\affiliation[f]{Leinweber Center for Theoretical Physics, University of Michigan, Ann Arbor, MI 48109, USA}

\emailAdd{sgori@ucsc.edu}
\emailAdd{zliuphys@umd.edu}
\emailAdd{bshakya@ucsc.edu}

\abstract{The electroweak sector of the Minimal Supersymmetric Standard Model (MSSM) -- neutralinos, charginos and sleptons -- remains relatively weakly constrained at the LHC due in part to the small production cross sections of these particles. In this paper, we study the prospects of searching for decays of heavy Higgs bosons into these superpartners at the high luminosity LHC. 
In addition to the kinematic handles offered by the presence of a resonant particle in the production chain, heavy Higgs decays can be the dominant production mode of these superpartners, making it possible to extend coverage to otherwise inaccessible regions of the supersymmetry and heavy Higgs parameter space. We illustrate our ideas with detailed collider analyses of two specific topologies: We propose search strategies for heavy Higgs decay to a pair of neutralinos, which can probe heavy Higgs bosons up to 1 TeV in the intermediate tan $\beta (\sim 2-8)$ region, where standard heavy Higgs searches have no reach. Similarly, we show that targeted searches for heavy Higgs decays into staus can probe stau masses up to several hundred GeV. We also provide a general  overview of additional decay channels that might be  accessible at the high luminosity LHC. This motivates a broader program for LHC heavy Higgs searches.
}

\preprint{
\begin{flushright}
FERMILAB-PUB-18-659-T
\end{flushright}
}

\maketitle

\section{Introduction}

There exists an amusing saying that half of the particles of the Minimal Supersymmetric Standard Model (MSSM) have already been discovered. This claim, however, is inaccurate: in addition to the hitherto undiscovered R-parity odd supersymmetric states that account for half of the MSSM particle content, the heavy Higgs bosons of the MSSM -- the scalar $H$, pseudoscalar $A$, and charged Higgs bosons $H^\pm$ -- also remain to be discovered. This amusing observation has important practical implications for the (R-parity conserving) MSSM: while the R-parity odd superpartners can only be pair-produced at the Large Hadron Collider (LHC), these heavy Higgs bosons can be singly produced. In addition to being advantageous from the energy viewpoint, such resonant production also enables additional kinematic handles for the final decay products, which can be tremendously useful in searching for signatures of such processes at the LHC and extending the coverage of the supersymmetry (SUSY) parameter space. This idea forms the central theme of this paper.

While the LHC provides stringent bounds on strongly interacting particles such as the gluino and squarks, significant gaps remain in the coverage of the electroweak sector of the MSSM, which consists of sleptons, charginos, neutralinos, and heavy Higgs bosons. For the sleptons, Higgsinos, and binos, the reach is relatively weak (searches and current limits are discussed in Section\,\ref{sec:framework}) due to their small direct pair-production cross sections. For the heavy Higgs bosons, the coverage is incomplete due to a lack of clean signatures at the LHC -- there exist strong limits at large tan\,$\beta \gtrsim 10$ due to the presence of the $A/H\to\tau\tau$ decay channel \cite{Aaboud:2017sjh,Sirunyan:2018zut}, but the reach at low tan\,$\beta\lesssim 8$ is relatively weak despite higher production cross sections since the dominant decay channel is $t\bar{t}$, which is a very challenging signal due to its interference with the SM $t\bar t$ background {\cite{Dicus:1994bm} (see \cite{Carena:2016npr,Gori:2016zto,Craig:2015jba,Jung:2015gta} for more recent studies). 

It is important to extend the LHC coverage for these electroweak SUSY particles, as they appear in several well-motivated MSSM frameworks. For instance, light neutralinos generically feature in well-tempered dark matter (DM) \footnote{In this paper, we do not consider dark matter constraints on the MSSM parameter space since the focus of the paper is on collider phenomenology, and consistency with cosmological constraints can be achieved in several ways that are unrelated to collider phenomenology. For instance, if the Lightest-Supersymmetric-Particle (LSP) is stable on collider scales but not on cosmological scales, as can occur for small R-parity breaking, then the collider discussions are unaffected (since the LSP still produces missing energy in the LHC detectors) but cosmological/ dark matter bounds are nonexistent (provided the LSP decays before BBN). Even with a cosmologically stable LSP, issues such as an excessive relic density or large direct detection cross sections can be successfully addressed in several different ways (e.g. entropy dilution from late decays of a heavy particle). Given such caveats, we do not consider dark matter constraints on our parameter space, as these would artificially constrain the available parameter space that is within reach of the LHC.} scenarios \cite{ArkaniHamed:2006mb} or within natural SUSY spectra~\cite{Papucci:2011wy}. Likewise, light staus are predicted in gauge-mediated SUSY breaking models \cite{Giudice:1998bp} and in DM models with DM-stau co-annihilation \cite{Ellis:1998kh,Ellis:1999mm,Han:2014nba}. Given the problems discussed above, significant enhancement of the reach for these particles at the LHC requires new production modes with higher cross sections, cleaner final states, or improved search strategies. To this end, in this paper we propose new LHC search strategies to broadly cover scenarios where these weakly interacting supersymmetric particles are produced from the decays of the heavy Higgs bosons, and we analyze their reach at the high luminosity (HL)- LHC. (Previous studies for heavy Higgs bosons decaying to electroweak particles can be found in \cite{Abdullin:2005yn,Moortgat:2001pp,Denegri:2001pn,Moortgat:2001wj,Ball:2007zza,Arganda:2012qp,Craig:2015jba,Barman:2016kgt,Kulkarni:2017xtf,Medina:2017bke,Arganda:2018hdn,Bahl:2018zmf,Han:2013gba,Baer:2013xua,Baer:2015tva}, and in Refs.~\cite{Dutta:2014hma,Christensen:2013dra,Han:2014nba,Ellwanger:2017skc,Baum:2017gbj,Baum:2019uzg} for Next-to-Minimal-Supersymmetric-Standard-Model studies.). In large regions of parameter space, such decays can account for the dominant production modes of these particles, with production cross sections significantly larger than those from direct production. In addition, the presence of the heavy Higgs resonance in the decay chain offers additional kinematical handles to identify these signal events over potentially large background. 

Following a general discussion of the content, interactions, LHC searches, and constraints in the electroweak sector of the MSSM in Section\,\ref{sec:framework}, we perform in-depth studies of two distinct topologies to illustrate the above ideas. First, in Section \ref{Sec:neutralinos}, we study the decay of the heavy Higgs bosons produced from gluon fusion into a heavier and an LSP neutralino, where the heavier neutralino subsequently decays into the LSP and a Z boson (see Fig.\,\ref{Fig:diagrams} (left)), yielding a $Z$+missing energy ($\met$) signal. Section \ref{Sec:staus} focuses on heavy Higgs bosons produced in association with b-quarks decaying into a pair of staus, where each stau decays into a tau lepton and the LSP neutralino (Fig.\,\ref{Fig:diagrams} (right)). Targeted search strategies making use of kinematic variables in these two scenarios\footnote{Note that there is also a recent ATLAS leptons plus $~\met$ excess that can be related to new exotic decays of the heavy Higgs bosons~\cite{Aaboud:2018sua,Carena:2018nlf,Lara:2018zvf}.} will be shown to significantly improve the reach for heavy Higgs bosons in the intermediate tan $\beta$ $(\sim2-8)$ region and for staus up to several hundred GeV respectively; our main results are presented in Figures \ref{Fig:chimatbreach} and \ref{Fig:staulimits}. Following these detailed studies, in Section \ref{Sec:others}, we offer an overview of additional promising signals that can be looked for at the LHC in the coming years. A summary of our results along with some concluding remarks are presented in Section \ref{Sec:summary}.

\begin{figure*}[t] 
\vspace{0.cm}
\includegraphics[width=6.9cm]{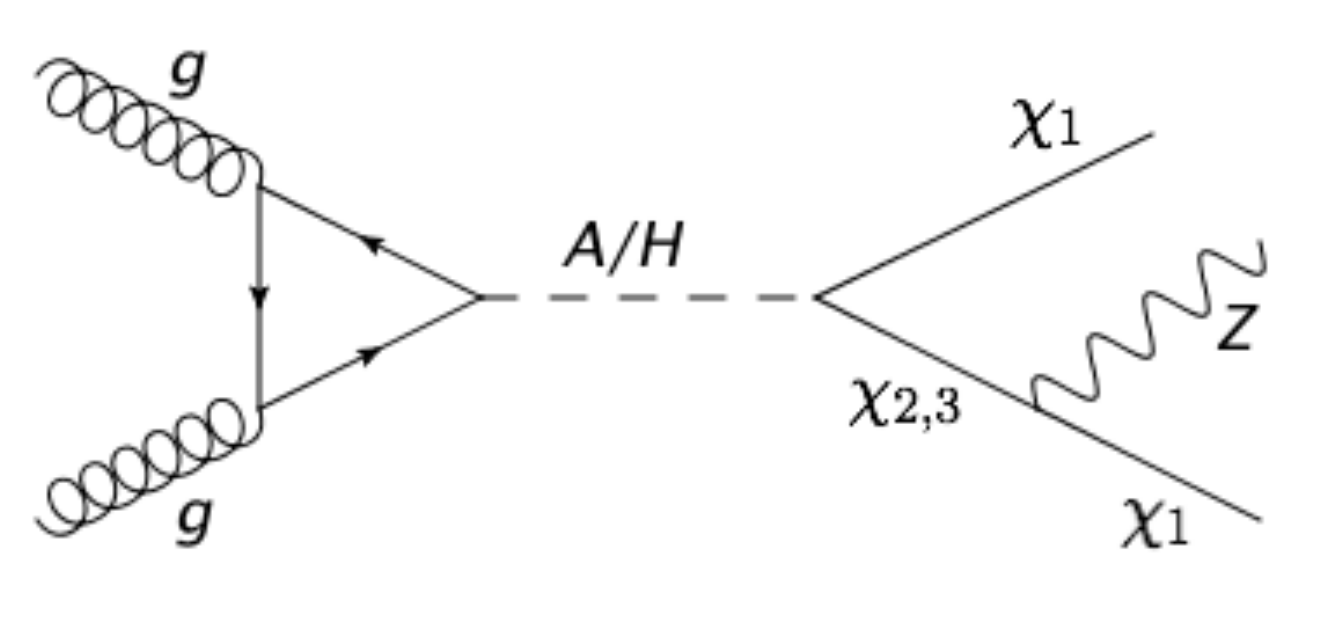}  ~~~~~~ \includegraphics[width=7cm]{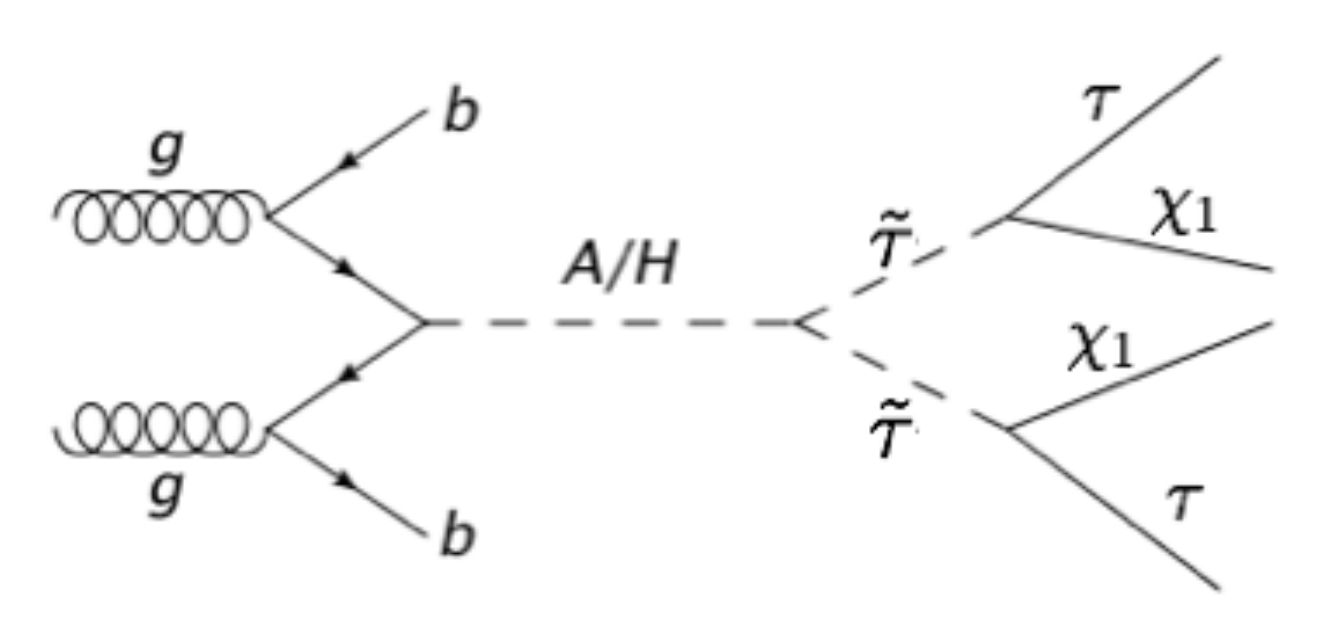}
\caption{Feynman diagrams for the two signal topologies we study in detail in this paper. {\bf{Left Panel}}: gluon fusion production of heavy Higgs bosons, which decay to neutralinos, yielding a $Z$+$\met$ signal. {\bf{Right Panel}}: b-associated production followed by decay to staus, giving a $2b+2\tau$+$\met$ signal.}
\label{Fig:diagrams}
\end{figure*}

\section{Framework: electroweak sector of the MSSM}
\label{sec:framework}

In this section, we discuss various experimental constraints on the electroweak sector of the MSSM, and examine various exotic decay channels of the heavy Higgs bosons and the relevant branching ratios over the parameter space of interest.

\subsection{Present LHC searches and constraints}\label{sec:constraintsEWinos}

The electroweak sector is the least constrained sector of supersymmetric models. In the MSSM in particular, current bounds on electroweakinos (neutralinos and charginos) from $\sim 36$ fb$^{-1}$ of 13 TeV data are at around 650 GeV \cite{Sirunyan:2018ubx, Aaboud:2018jiw} in the most collider-favorable scenarios, where the next-to-lightest supersymmetric particle (NLSP) is wino-like and the bino-like LSP is massless. Bounds weaken to $\sim 450$ GeV for Higgsino-like NLSPs due to their smaller pair production cross sections. In contrast, current bounds for SUSY particles produced through strong interactions are at the level of $\sim$ 2 TeV (gluinos) and $\sim$ 1.5 TeV (squarks) ~\cite{Aaboud:2017vwy, Sirunyan:2017kqq}.

The main electroweakino signatures driving the $\sim650$ GeV bounds are $pp\to\chi^\pm \chi_2,~\chi^\pm\to\chi_1 W^{(*)},\,\chi_2\to\chi_1 Z^{(*)}$, with the $W$ and $Z$ bosons producing leptons or jets in the final state, resulting in $3\ell+\,\met$ and $2\ell+$jets+$\met$ signatures. Additionally, searches for $\ell+2b+\met$ \cite{Sirunyan:2017zss} have been performed to set constraints on the decay topology $pp\to\chi^\pm \chi_2,~\chi^\pm\to\chi_1 W^{(*)},~\chi_2\to\chi_1 h^{(*)}$, constraining wino masses up to $\sim 500$ GeV for massless LSPs. Finally, direct searches for neutralino and chargino-pair production have been also performed in the $4\ell+\met$ ($pp\to\chi_2 \chi_2,~\chi_2\to\chi_1 Z^{(*)}$) \cite{Aaboud:2018zeb} and $2\ell+\met$ ($pp\to\chi^\pm \chi^\mp,~\chi^\pm\to\chi_1 W^{(*)}$) channels \cite{Sirunyan:2018lul,Sirunyan:2018ubx, Aaboud:2018jiw}. These searches from neutral current DY process are less sensitive due to either lower cross sections ($4\ell+~\met$) or larger backgrounds ($2\ell+~\met$), and probe $\sim 200$ GeV wino-like charginos for massless LSP.

Bounds are significantly weaker for electroweakino spectra with massive LSPs due to smaller amounts of missing energy as well as smaller $p_T$ of the visible objects in the final state. For example, for a mass splitting of 100 GeV between NLSP and LSP, the bound on wino-pair production is $\sim230$ GeV ~\cite{Sirunyan:2018ubx}. The corresponding exclusion for Higgsino-pair production, derived from the upper limit on the cross section from this search, is at $\sim 180$ GeV. For $m_{LSP}> 250\,(150)$ GeV, the bounds essentially vanish for wino (Higgsino) NLSP. 

For models with slepton NLSP, there are comparable constraints of around 500 GeV for massless LSPs \cite{Aaboud:2018jiw,Sirunyan:2018nwe} from searches for pair produced sleptons with multilepton final states $pp\to\tilde\ell\tilde\ell,~\tilde\ell\to \ell+\met$. 
  An exception to this are third generation sleptons (ie. staus), for which  the Drell-Yan (DY) stau pair production cross section is relatively small, and the corresponding signature, $pp\to \tilde\tau\tilde\tau\to 2 \tau+\met$, is background limited due to the large irreducible $pp\to Z Z/\gamma^*,~W^+W^-$ backgrounds, resulting in LHC bounds \cite{Sirunyan:2018vig} that are still weaker than the corresponding LEP bounds \cite{Abdallah:2003xe} (for additional channels that could constrain the stau parameter space, see \cite{Carena:2012gp}). 

Given that direct production bounds on the wino are relatively strong, in this paper we decouple the wino from our analysis and focus on scenarios with light Higgsinos and bino. Likewise, since the focus of this paper is on probing electroweak particles from heavy Higgs boson decays, and since their couplings to sleptons are proportional to the corresponding lepton masses, we will decouple the first and second generation sleptons and focus on the third generation sleptons, staus.
As we will demonstrate in the following sections, Higgsinos and staus can have sizable interactions with the Higgs sector of the MSSM, producing interesting exotic signatures for heavy Higgs boson decays. Our discussion can easily be extended to the Next-to-Minimal-Supersymmetric-Standard-Model (NMSSM), where the LSP can be the singlino instead of the bino (however, additional structure in the NMSSM can also lead to other signals and decay topologies).

\subsection{Higgs-electroweak sector interactions}

We follow the notation and conventions of Ref.\,\cite{Martin:1997ns}, and take the Higgs vacuum expectation value (vev) $v=174$ GeV.
In addition to the 125 GeV Higgs boson, $h$, the Higgs sector of the MSSM consists of a CP-even scalar, $H$, a CP-odd pseudoscalar, $A$, and a charged Higgs, $H^{\pm}$. To simplify the discussion, we consider no new sources of CP violation. In this scenario, the above Higgs bosons are distinct mass eigenstates, with $m_H\approx m_A$ and $m_{H^{\pm}}^2=m_A^2+m_W^2$. Below, we review the interactions of these states with the electroweakinos and sleptons. We will only discuss the interactions that will be relevant to our paper; extended reviews of the MSSM interactions can be found in, e.g., \cite{Djouadi:2005gj}.

The neutralino sector consists of the bino $\tilde{B}$, wino $\tilde{W}$, and the two Higgsinos $\tilde{H}_d, \tilde{H}_u$. The neutralino mass matrix in this basis is given by\,\footnote{Dots in the mass matrices indicate entries of a symmetric matrix.}
\begin{eqnarray}
	M_{\chi}= \begin{pmatrix} M_1 & 0 & -m_Z\, s_W\,c_\beta & m_Z\,s_W\,s_\beta 
	\\ \cdot & M_2 & m_Z\,c_W\,c_\beta & -m_Z\,c_W\,s_\beta 
	\\ \cdot& \cdot& 0 & -\mu 
	\\ \cdot& \cdot& \cdot& 0 
 \end{pmatrix},
\end{eqnarray}
where we have denoted $s_W\equiv \sin\theta_W$, with $\theta_W$ the weak mixing angle, $s_\beta\equiv \sin\beta$ and similar for $c_W,c_\beta$. The mass eigenstates can be written as 
\beq
\tilde{\chi_i}=Z_{i1}\tilde{B}+Z_{i2}\tilde{W}+Z_{i3}\tilde{H}_d+
Z_{i4}\tilde{H}_u\, ,
\eeq
with $i=1-4$ labeling the mass eigenstates from lightest to heaviest. We decouple the wino as discussed above, thus $Z_{i2}\approx 0$ for $i=1,2,3$. Similarly, the chargino sector consists of a charged wino and a charged Higgsino with a mass matrix in this basis given by
\begin{eqnarray}
	M_{\chi^\pm}= \begin{pmatrix} M_2 & \sqrt 2 \,m_W\, s_\beta\\ 
	\sqrt 2 \,m_W \,c_\beta & \mu 
 \end{pmatrix}.
\end{eqnarray}
Decoupling the wino simply leaves Higgsino-like chargino states $\chi^\pm$, with $m_{\chi^\pm}\sim\mu$.

Finally, the slepton sector is composed of three generations of sleptons. The mass matrix for the third generation, which we focus on, is given by
\begin{eqnarray}
	M^2_{\tilde\tau}= \begin{pmatrix} m^2_{\tilde\tau_L} + m_\tau^2 +c_{2\beta} \,m_Z^2(s_W^2-1/2)& m_\tau (A_\tau-\mu\, t_\beta) \\ 
	 \cdot & m^2_{\tilde\tau_R}+ m_\tau^2 -c_{2\beta} \,m_Z^2 \,s_W^2
 \end{pmatrix},
 \label{eq:staumass}
\end{eqnarray}
where $A_\tau$ is the dimensionful trilinear coupling from the soft term, $y_\tau A_\tau H_d \tilde{\bar\tau}_L \tilde{\tau}_R$. 

The Higgs bosons couple to the bino-Higgsino and wino-Higgsino combinations. In the decoupling/alignment limit \cite{Gunion:2002zf} (for more recent studies, see \cite{Delgado:2013zfa,Craig:2013hca,Carena:2013ooa,Haber:2013mia}), where the lighter Higgs boson, $h$, is SM-like and the heavier scalar, $H$, is the orthogonal component, the couplings between the heavy Higgs bosons ($H,A$) and neutralinos (with the wino decoupled) are (see e.g.\cite{Djouadi:2001kba}) 
\bea
g_{H\chi_i\chi_j}&=&\frac{g'}{2}Z_{i1}(Z_{j3}\,\sin\beta + Z_{j4}\,\cos\beta  )+(i \leftrightarrow j)\nonumber\\
~~g_{A\chi_i\chi_j}&=&\frac{g'}{2}Z_{i1}( Z_{j3}\,\sin\beta -Z_{j4}\,\cos\beta)+(i \leftrightarrow j),
\label{eq:Hchichi}
\eea
for $i,j=1,2,3$. Here $g'^2=2 m_Z^2 s_W^2/v^2$. In our framework, where the relevant light chargino is purely Higgsino, the charged Higgs coupling to a chargino-neutralino pair is particularly simple, $|g_{H^+\chi^-\chi_i}|=\frac{1}{\sqrt{2}}g'Z_{i1}$, whereas the coupling of the neutral Higgs bosons to charginos vanishes.

For staus, the couplings to the heavy scalar and pseudoscalar Higgs bosons in the decoupling/alignment limit are given by
\begin{eqnarray}
&&|g_{H \tilde{\tau}_L \tilde{\tau}_L}|=\sqrt{2} \left|s_{2\beta} \frac{m_Z^2}{v}(s_W^2-\frac{1}{2})-\frac{m_\tau^2}{v}\right|, ~~|g_{H \tilde{\tau}_R \tilde{\tau}_R}|=\sqrt{2} \left|s_W^2\,s_{2\beta} \frac{m_Z^2}{v}+\frac{m_\tau^2}{v}\right|,\\
&& |g_{H \tilde{\tau}_L \tilde{\tau}_R}|=\frac{m_\tau}{\sqrt{2}v} |A_\tau\, \tan\beta+\mu|,~~ |g_{A \tilde{\tau}_L \tilde{\tau}_R}|=\frac{m_\tau}{\sqrt{2}v} |A_\tau\, \tan\beta-\mu|.
\label{eq:staucouplings}
\end{eqnarray}
For the sneutrinos, the most relevant coupling is to the charged Higgs and a stau:
\begin{equation}\label{eq:staucouplingsPM}
|g_{H^\pm \tilde{\nu}_\tau\tilde{\tau}_R}|=\frac{m_\tau}{v} |A_\tau\, \tan\beta-\mu|,~~~|g_{H^\pm \tilde{\nu}_\tau\tilde{\tau}_L}|=\left|\frac{m_W^2}{v} \sin 2\beta -\frac{m_\tau^2}{v}\tan \beta\right|.
\end{equation}

\subsection{Higgs branching ratios to supersymmetric electroweak particles}\label{sec:2.3}

As discussed above, we decouple all supersymmetric electroweak particles except the bino, Higgsinos, and the third generation sleptons. Several exotic decay channels of the heavy Higgs bosons into these states are possible:
\beq
A/H\to \chi_i \chi_j,\,\chi^+ \chi^-,\, \tilde{\tau}_i\tilde{\tau}_j, \,\tilde{\nu}\tilde{\nu},~~~~~H^\pm\to \chi_i\chi^\pm,\,\tilde{\tau}\tilde{\nu}.
\eeq 
In Fig.\,\ref{Fig:branchingratios}, we plot the branching ratios for these final states as a function of tan\,$\beta$ for an illustrative benchmark scenario with $m_A=800$ GeV, $m_{\tilde{\tau}_L}=m_{\tilde{\tau}_R}=\mu=350$ GeV, $M_1$=150 GeV, $A_{\tau}=1$ TeV, and all other dimensionful parameters set to $2$ TeV. We only show channels with branching ratios above $1\%$. We see that several channels can have $\mathcal{O}(10)\%$ branching ratios, which depend non-trivially on tan\,$\beta$. 

\begin{figure*}[t] 
\vspace{0.cm}
\includegraphics[width=5.1cm]{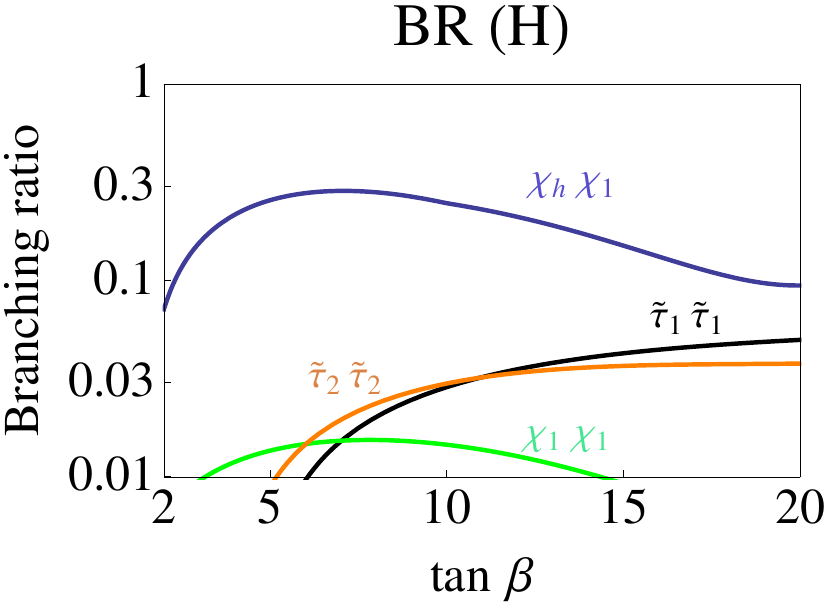}  \includegraphics[width=5.1cm]{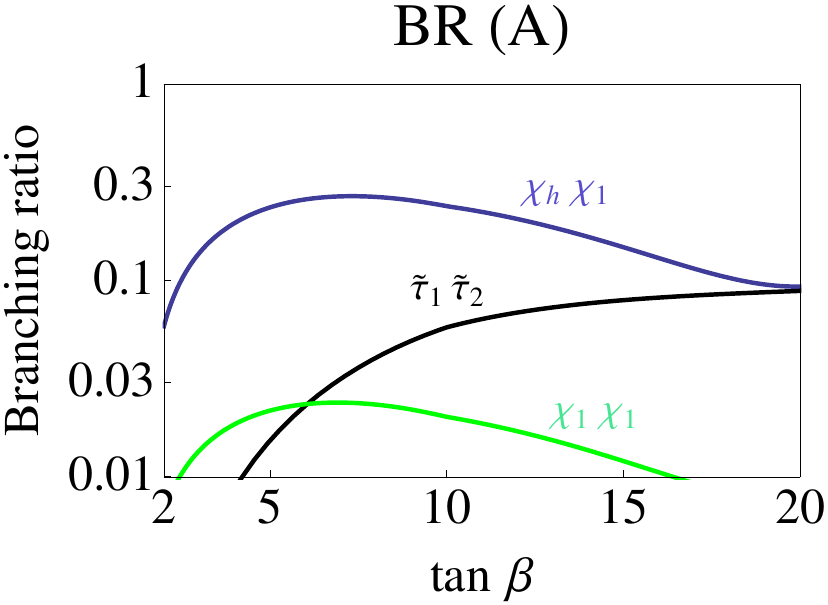} \includegraphics[width=5.1cm]{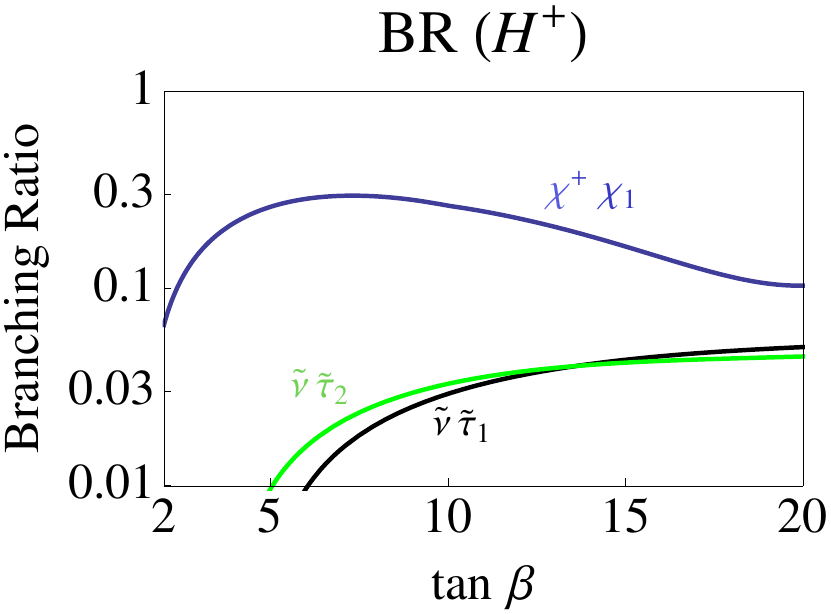}
\caption{Branching ratios of the heavy Higgs bosons into the various electroweak states, as calculated with {\tt FeynHiggs} 2.10.2 \cite{Heinemeyer:1998yj}. For these plots, we set $m_A=800$ GeV, $m_{\tilde{\tau}_L}=m_{\tilde{\tau}_R}=\mu=350$ GeV, $M_1$=150 GeV, $A_{\tau}=1$ TeV, and all other dimensionful parameters to $2$ TeV. $\chi_h\chi_1$ denotes the sum over $\chi_2\chi_1$ and $\chi_3\chi_1$. We only show channels with branching ratios above $1\%$.}
\label{Fig:branchingratios}
\end{figure*}

The heavy Higgs bosons couple to gaugino-Higgsino combinations in neutralinos and charginos (see Eq.\,\ref{eq:Hchichi}). For our choice of $M_1=150$ GeV and $\mu=350$ GeV, $\chi_1$ is bino-like and $\chi_{2,3}$ are Higgsino-like, with small ($\sim m_Z/\mu\sim 0.25$) mixing. This results in large branching ratios for $A/H$ couplings to neutralino combinations that are bino-Higgsino like\,\footnote{Such mass spectra and couplings are also expected in scenarios with neutralino dark matter compatible with indirect and direct detection constraints, see e.g. discussions in \cite{Profumo:2017ntc,Perelstein:2012qg,Perelstein:2011tg}.}, i.e. $\chi_3\chi_1$ and $\chi_2\chi_1$ (see blue curves in the left and center panels in Fig. \ref{Fig:branchingratios}), whereas the branching ratios into the remaining $\chi_i\chi_j$ neutralino pair combinations, with $\{i,j\}=\{1,1\},\{2,2\},\{2,3\},\{3,3\}$, are suppressed by this small mixing angle. For similar reasons, the charged Higgs branching ratio to the Higgsino-bino combination $\chi^+\chi_1$ is unsuppressed, but the combinations $\chi^+\chi_2$ and $\chi^+\chi_3$ are again suppressed by this mixing angle, leading to branching ratios below the percent level (right panel of the figure). 
These electroweakino couplings do not depend strongly on the value of tan\,$\beta$, hence the corresponding branching ratios peak at tan$\beta\sim 7$, where the total $A/H$ width is minimized as neither the up-type nor down-type Yukawas are too large, as seen in the plot.

In contrast, the heavy Higgs couplings to the third generation sleptons are proportional to tan\,$\beta$ for $A_{\tau}$\,tan\,$\beta\gg\mu$ (see Eq.\ref{eq:staucouplings},\,\ref{eq:staucouplingsPM}), and the relevant curves in Fig.\,\ref{Fig:branchingratios} show that the corresponding branching ratios increase accordingly at higher tan\,$\beta$. Note that, unlike the neutral scalar $H$, the pseudoscalar $A$ cannot  decay into identical pairs $\tilde{\tau}_1\tilde{\tau}_1$ or $\tilde{\tau}_2\tilde{\tau}_2$ because of CP conservation, and decays instead to $\tilde{\tau}_1\tilde{\tau}_2$.

Based on these observations, the $\textit{unsuppressed}$ decay channels, which are most promising for collider searches, are
\beq
A/H\to \chi_{(2,3)} \chi_1,\, \tilde{\tau}_i\tilde{\tau}_j,~~~~~H^\pm\to \chi_1\chi^\pm,\,\tilde{\tau}\tilde{\nu}.
\eeq 
Indeed, both previously studied \cite{Carena:2013ytb} and recent \cite{Bahl:2018zmf} MSSM benchmark scenarios used to interpret LHC searches for SUSY heavy Higgs bosons predict sizable branching ratios of $A/H$ into either neutralinos/charginos ($m_h^{\rm{max}}$, $m_h^{\rm{mod}\pm}$ scenarios) or staus ($\tau-$phobic, light stau scenarios). We will therefore focus on these channels in the remainder of this paper. In particular, we will perform in-depth collider studies for $A/H\to \chi_{(2,3)} \chi_1,\, \tilde{\tau}_i\tilde{\tau}_j$, and, in Sec.\,\ref{Sec:others}, we will discuss the prospects and benchmarks for the charged Higgs decays, $H^\pm\to \chi_1\chi^\pm,\,\tilde{\tau}\tilde{\nu}$.

\subsection{LHC rates for electroweak production through Higgs decays}\label{sec:2.1}

In Fig.\,\ref{Fig:Branchingratios}, we show the branching ratios Br$(A\to$ neutralinos) and Br$(A\to$ staus), calculated using the package {\tt SUSY-HIT} \cite{Djouadi:2006bz}, as a function of the heavy Higgs mass $m_A=m_H$ and $\tan\beta$.\footnote{Depending on the exact SUSY spectrum, sizable electroweak corrections could appear for these calculations, see Ref.~\cite{Heinemeyer:2014yya,Heinemeyer:2015pfa}.} We use a benchmark with $M_1=150$ GeV, $\mu=m_A-175$ GeV, $M_2=2$ TeV, $A_f=\mu\,/\,$tan$\,\beta + 1600$ GeV, and all other dimensionful parameters fixed to $1.5$ TeV.
Br$(H\to$ neutralinos, staus) are not shown in these figures as they are numerically similar to the pseudoscalar ones.
We see that, over large regions of parameter space not ruled out by standard $H\to\tau\tau$ searches\,\footnote{Here, and in the remainder of the paper, we take the stronger of the ATLAS or CMS $H\to\tau\tau$ limits \cite{Aaboud:2017sjh,Sirunyan:2018zut}, considering both gluon fusion and b-associated production processes.}, the heavy Higgs bosons can have $\mathcal{O}(10\%)$ branching ratios into these electroweak states, suggesting that these final states could provide observable channels at colliders. Furthermore, as shown in the color coding in these plots, the production cross section of these electroweak states via heavy Higgs decays can be substantially larger (by over an order of magnitude in some cases) than the corresponding direct production cross section\,\footnote{In calculating the several branching ratios and production cross sections, we have ignored loop contributions from SUSY states, which can  significantly affect the $(A/H)~ b\bar b$ and $(A/H)~ \tau\bar\tau$ couplings \cite{Hall:1993gn,Hempfling:1993kv,Carena:1994bv,Pierce:1996zz,Carena:2012rw}. We have checked that these lead to percent level modifications of the $Ab\bar b, A\tau\tau$ couplings for the parameters we consider, which can be further suppressed by raising some parameter (such as $M_2$ or sfermion masses, which do not affect the decay channels we are interested in) to higher values.}. Hence heavy Higgs decays could provide the dominant source of production of these electroweak particles, offering opportunities to detect these particles over direct production.

\begin{figure*}[t] 
 \vspace{0.cm}
  \includegraphics[width=8.25cm]{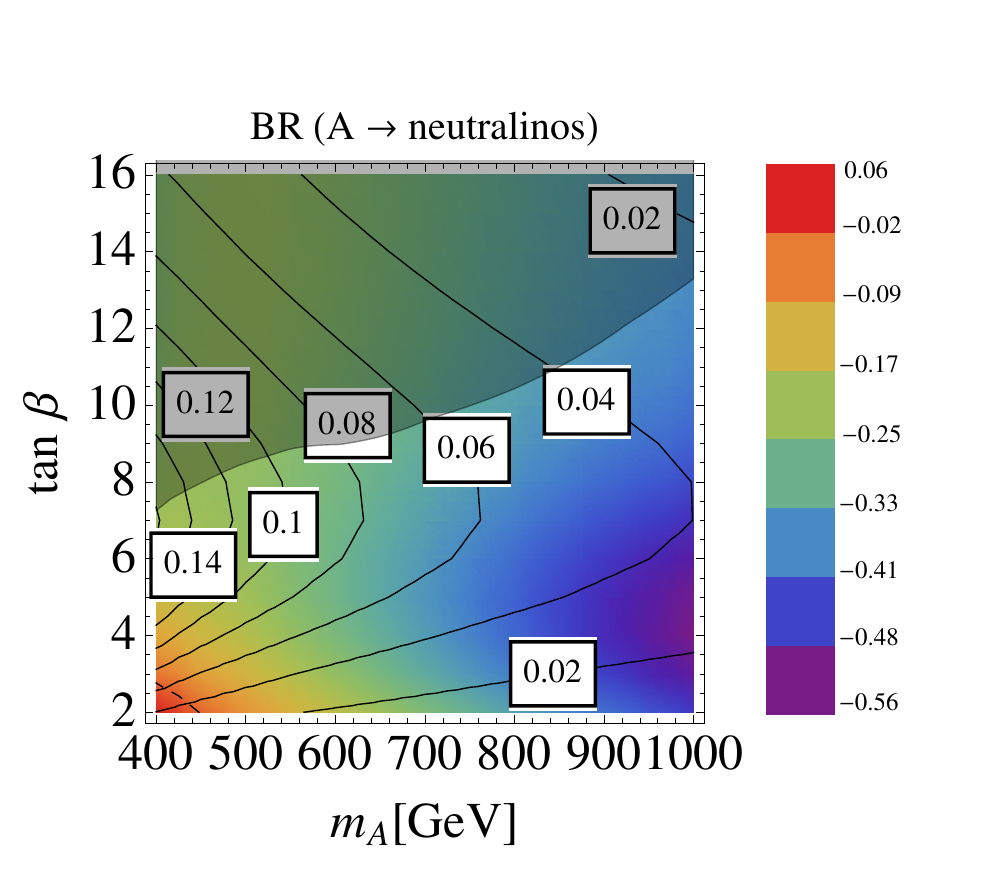}  \includegraphics[width=8.2cm]{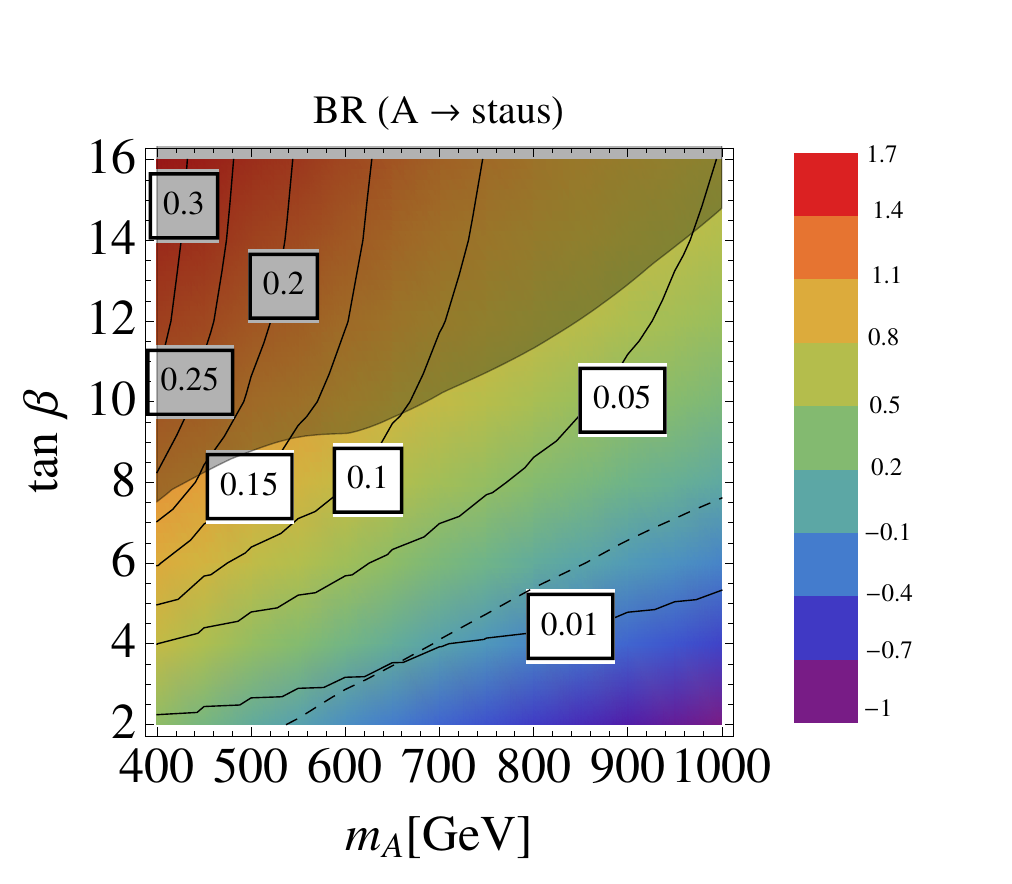}
\caption{{\bf Left Panel:} Curves with numerical labels denote contours of Br$(A\to$ neutralinos). For this scan, we set $M_1=150$ GeV, $\mu=m_A-175$ GeV, $M_2=2$ TeV, all soft terms $A_f=\mu\,/\,$tan$\,\beta + 1600$ GeV, and all other dimensionful parameters to $1.5$ TeV. These parameters are similar to the $m_h^{\rm{mod}+}$ scenarios from \cite{Carena:2013ytb}, except for the modified values of $M_1, M_2,$ and $\mu$. The color coding represents the ratio of electroweakino production cross section via heavy Higgs decay to direct Drell-Yan production cross section, log$_{10}\,(\sigma(pp\to A,H\to$ neutralinos)/$\sigma(pp\to$ electroweakinos)), where the numerator includes contributions from both gluon fusion and b-associated production of $A$ and $H$, and the denominator includes all direct neutralino and chargino production modes. The dashed line indicates the contour along which these two production cross sections are equal. {\bf{Right Panel:}} Analogous plot for Br$(A\to$ staus), with parameter choices the same as for the neutralino plot, except $m_{\tilde{\tau}_L}=m_{\tilde{\tau}_R}=m_A/2 -50$ GeV, $A_{\tau}=1$ TeV, and $\mu=500$ GeV, which are similar to the light stau scenario in \cite{Carena:2013ytb} except for the modified values of $A_\tau, m_{\tilde{\tau}_L}, m_{\tilde{\tau}_R},$ and $M_2$.  In both panels, the shaded regions represent the parameter space excluded by currents LHC searches for $A/H\to\tau\tau$ \cite{Aaboud:2017sjh,Sirunyan:2018zut}, whereas all non-shaded regions are consistent with all current LHC bounds from various searches.}
\label{Fig:Branchingratios}
\end{figure*}

The two panels in the plot reveal contrasting behaviors, both in terms of production mechanism and of the branching ratios, for the two decay channels of interest. As mentioned earlier, the heavy Higgs boson couplings to SM fermions are smallest at intermediate values of tan\,$\beta\sim 7$, where neither the up-type ($\sim m_t/\tan\beta /v$) nor down-type ($\sim m_b\tan\beta /v$) fermion couplings are too large. We see that the largest values of Br$(A\to$ neutralinos) are realized in this regime (left panel). In contrast, the heavy Higgs couplings to staus are proportional to tan\,$\beta$ if $A_\tau \tan\beta/\mu>1$ (see Eq.\,\ref{eq:staucouplings}), hence Br$(A\to$ staus) grows with tan\,$\beta$ (right panel). For sufficiently large tan\,$\beta$, however, the dominant decay is into bottom quarks, which scales similarly with tan\,$\beta$, hence Br$(A\to$ staus) approaches a constant value. 

The dominant production mechanism for heavy Higgs bosons has important bearing on the optimal parameter space and search strategy at the LHC. Their production is dominated by gluon fusion at low values of tan\,$\beta$ and $b-$associated production at high values of tan\,$\beta$. For the decay into neutralinos, for which the couplings to Higgs bosons do not depend strongly on tan\,$\beta$, we see that the most promising regime for producing a strong signal compared to direct electroweak production is at low tan\,$\beta$, where gluon fusion leads to a sizable heavy Higgs production cross section thanks to the top loop, while BR($A\to$ neutralinos) remains sizable (see the red region in the plot). For the decay into staus, on the other hand, Br$(A\to$ staus) drops sufficiently rapidly at low tan\,$\beta$ that the sizable gluon fusion production cross section is no longer relevant; the optimal region of parameter space lies instead at large tan\,$\beta$, where both the $b-$associated production cross section as well as BR($A\to$ staus) get enhanced. These promising scenarios for neutralinos and staus will be studied in detail in Sections \ref{Sec:neutralinos} and \ref{Sec:staus}, respectively.

\section {Searching for heavy Higgs decays to neutralinos} 
\label{Sec:neutralinos}

In this section, we study the prospects for probing heavy Higgs decays to neutralinos at the HL- LHC. 
In particular, we focus on the topology 
\beq
pp\to A,H \to \chi_1\chi_h,\,\chi_h\to \chi_1 Z,~~h=2,3,\eeq
with $A,H$ produced from gluon fusion\footnote{The inclusion of b-associated production will slightly improve our reach in some parts of parameter space; in this sense, our analysis is conservative.}, and where the LSP, $\chi_1$, is a bino-like neutralino ($m_{\chi_1}\sim M_1$), and the heavier neutralino states $\chi_h=\chi_{2,3}$ are Higgsino-like ($m_{\chi_3}\approx m_{\chi_2}\sim\mu$). To optimize the reach for the decay chain of interest, we consider the benchmark model $m_{\chi_3}\approx m_{\chi_2}\approx m_{\chi_1}+100$ GeV\,\footnote{We leave for future work the study of more squeezed scenarios with $m_{\chi_{2,3}}-m_{m_{\chi_1}} < m_Z$.},
so that $\chi_h\to \chi_1 h$ is kinematically forbidden and the heavier neutralinos decay exclusively as $\chi_h\to \chi_1 Z$, leading to a mono-$Z+\met$ signature.

 While direct production of $\chi_1\chi_h$ through Drell-Yan processes also leads to the same signature, the present bounds from standard $Z+\met$ searches \cite{ATLAS-CONF-2016-056,Aaboud:2016qgg,CMS-PAS-EXO-16-038,Aaboud:2017rel,Aaboud:2017bja,Sirunyan:2017hci,Sirunyan:2017jix} are relatively weak due to the relatively small mass splitting between the neutralino states resulting in a soft $Z$ (we will discuss this further in Sec. \ref{sec:colliderNeutralinos}). In contrast, in our setup the heavy Higgs decay kinematics provides additional boost to $\chi_h$, providing more energetic visible products even in this relatively compressed scenario.
One can also interpret mono-$Z$ searches in terms of the process $pp\to (\chi_1\chi_1) + Z$ arising in our benchmark models, with the $Z$ emitted as initial state radiation. This also provides a weak constraint due to the suppressed direct production cross section of a pair of (mostly bino) LSPs. As we will see below, the strongest constraints on this electroweak benchmark scenario come from multi-lepton searches for $pp\to \chi_{2,3}\chi_1^{\pm}\to \chi_1\chi_1 Z W $\cite{Aaboud:2018sua,Aaboud:2018jiw,CMS-PAS-SUS-16-034, CMS-PAS-SUS-16-039,CMS-PAS-SUS-17-004}, for which the bound on wino-pair production is $\sim230$ GeV for a 100 GeV mass splitting between NLSP and LSP, and the corresponding exclusion for Higgsino-pair production is $\sim 180$ GeV.

\subsection{Benchmark models}

As discussed in Sec.\,\ref{sec:2.1}, the optimal region of parameter space for heavy Higgs decay to neutralinos is at low tan\,$\beta$. In this regime, the gluon fusion production cross section of the heavy Higgs bosons is sizable, while BR($A/H\to$ neutralinos$)=\mathcal O(0.1)$, leading to the maximal production of neutralino states (also in comparison to direct Drell-Yan production of neutralinos). 

The contours in Fig.\,\ref{Fig:gluonfusion} show the production cross section for the process $pp\to A,H\to \chi_h \chi_1$ as a function of $m_A$ and $m_{\chi_3}$ for several values of tan\,$\beta$. We choose the parameter $M_1$ as a function of $\mu$ such that $m_{\chi_{3}}\approx m_{\chi_2}\approx m_{\chi_1}+100$ GeV \footnote{This mass splitting is mainly chosen for convenience, as with this spectrum, $\chi_h \rightarrow \chi_1 Z$ is the only available decay channel. The search strategy discussed below remains applicable for larger mass splittings; however, in this case $\chi_h \rightarrow \chi_1 h$ also opens up, suppressing the branching fraction into the $\chi_1 Z$ channel.}. We see that the production cross section for the process  ranges from $1-70$ fb for tan\,$\beta$ in the range $2-8$, with larger cross sections corresponding to lower values of tan\,$\beta$. The several colors in the plots correspond to the ratio of the cross section for this process to direct neutralino production cross section, $\sigma(pp\to A,H \to \chi_1(\chi_h\to \chi_1 Z))/\sigma(pp\to$ neutralinos$\to Z+\met$), where we sum over all possible combinations of $\chi_{1,2,3}$ in the denominator that can give rise to the $Z+\met$ signal. We see that the production from Higgs decay can be the dominant production mode for neutralinos giving rise to the $Z+\met$ signal in part of the parameter space (regions to the left of the dashed curves in the left and center panels). As we will see in the following subsection, even in regions where the direct DY production is larger, the presence of an on-shell heavy Higgs in the chain provides a crucial handle that can allow to probe these regions of parameter space beyond the reach of direct neutralino searches.

\begin{figure*}[t] 
\vspace{0.cm}
\includegraphics[width=5.07cm]{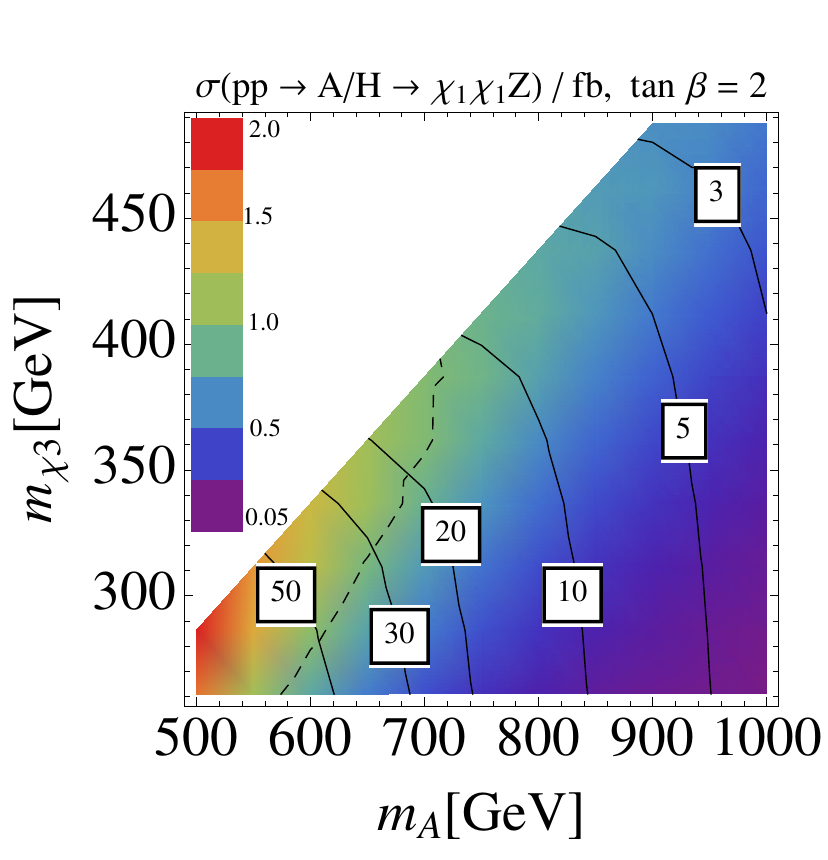}  \includegraphics[width=5.1cm]{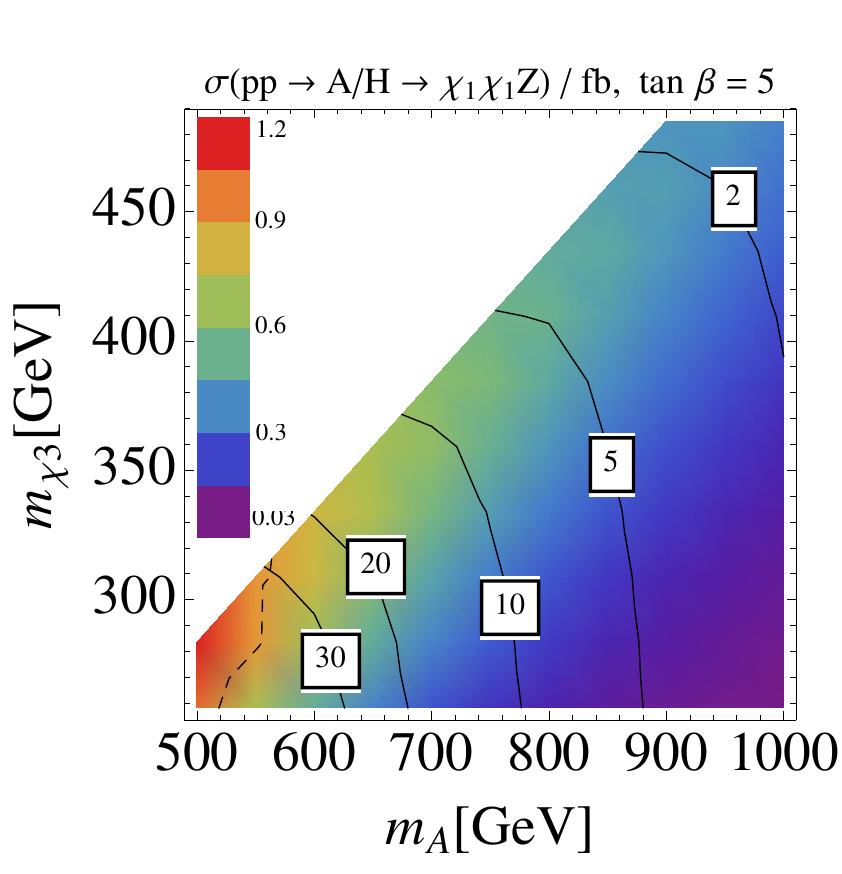} \includegraphics[width=5.1cm]{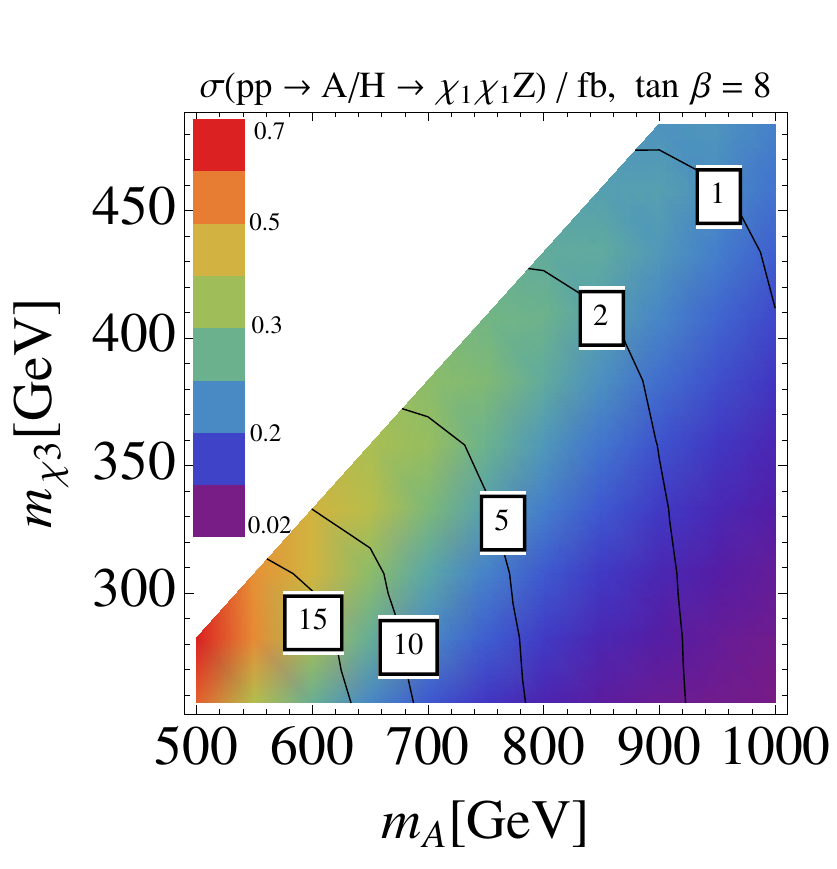}
\caption{ 13 TeV production cross section for the process $pp\to A,H \to \chi_1(\chi_h\to \chi_1 Z)$, where $h=2,3$ and we sum over both $A$ and $H$, as a function of $m_A$ and $m_{\chi_{3}}$ for $\tan\beta=2,5,8$ (left, center, and right panels, respectively). For these plots we fix $M_1$ such that $m_{\chi_{3}}\approx m_{\chi_2}\approx m_{\chi_1}+100$ GeV, and decouple all other particles by setting other dimensionful parameters to 2 TeV. The color coding represents the ratio of the cross section for this process to direct neutralino production cross section, $\sigma(pp\to A,H \to \chi_1(\chi_h\to \chi_1 Z))/\sigma(pp\to$ neutralinos$\to Z+\met$)), where we sum over all possible neutralino combinations that can give rise to a mono-Z signal in the denominator (including topologies such as $\chi_2\chi_3$, which gives a mono-Z signal when one of the $Z$s decays invisibly). The dashed curve in the left panel indicates the contour along which these two production cross sections (direct and via heavy Higgs decays) are equal. All regions of parameter space presented in these panels are consistent with current LHC bounds from $A/H\to\tau\tau$ \cite{Aaboud:2017sjh,Sirunyan:2018zut} as well as $pp\to \chi_2\chi_1^{\pm}\to \chi_1\chi_1 Z W$\,\cite{Aaboud:2018sua,Aaboud:2018jiw,CMS-PAS-SUS-16-034, CMS-PAS-SUS-16-039,CMS-PAS-SUS-17-004}.}
\label{Fig:gluonfusion}
\end{figure*}

\subsection{Proposed search} 
\label{sec:colliderNeutralinos}

We now turn to a Monte Carlo analysis of the signal of interest as well as the relevant backgrounds. We perform our numerical study using the leading order {\tt Madgraph5}~\cite{Alwall:2011uj} signal and background events showered through {\tt Pythia6}~\cite{Sjostrand:2006za}. The events are then passed through a fast detector simulation using {\tt Delphes3}~\cite{deFavereau:2013fsa} with the default {\tt Madgraph5} card.

The major SM background is from $Z$-boson pair production, with one $Z$ boson decaying leptonically and the other decaying invisibly into neutrinos, $pp\to ZZ \to \ell^+\ell^- +\met$. A subleading background contribution comes from $W$-boson pair production, with both $W$ bosons decaying leptonically to give a pair of same-flavor-opposite-sign dileptons. For the analysis below, we have checked that the $t\bar t+$ jets background is negligible. The SM background cross section at 13 TeV LHC into $\ell^+\ell^-+\met$ is 1.07~pb (with a minimal cut on the charged lepton $p_T$ of 10~GeV and maximal rapidity cut of $2.5$ at parton level). To compute this rate, we have applied a plain k-factor of 1.6~\cite{Grazzini:2015hta}. To reconstruct the $Z$ boson present in the decay chain, we impose that the dilepton pair is reconstructed near the $Z$-pole, $85~\gev < m_{\ell\ell} < 95~\gev$, which reduces the background to 0.20~pb, consisting mostly of $ZZ$ events.  

\begin{figure*}[t] 
  \vspace{0.cm}
   \center{
     \includegraphics[width=7.5cm]{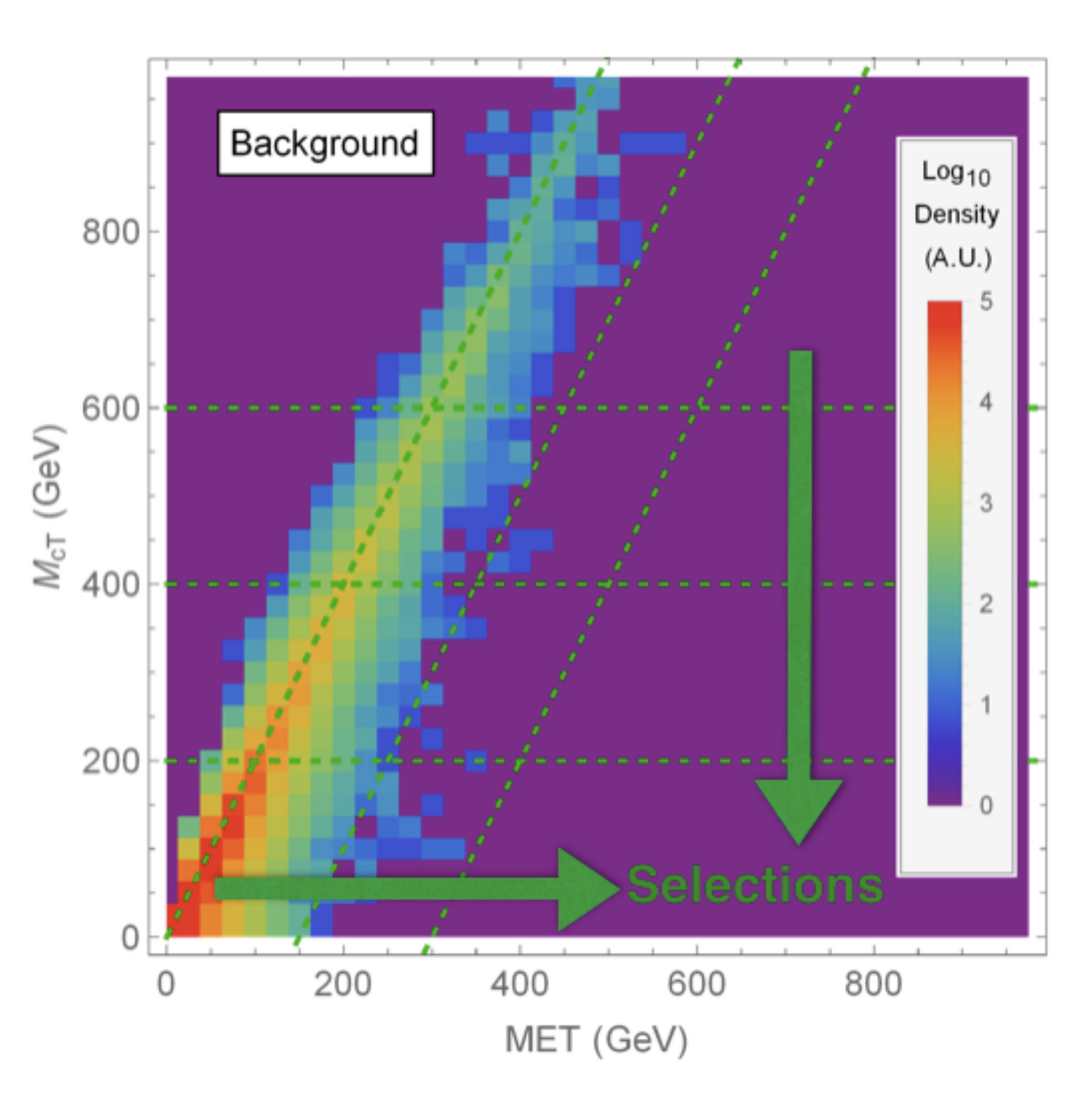}
     \includegraphics[width=7.5cm]{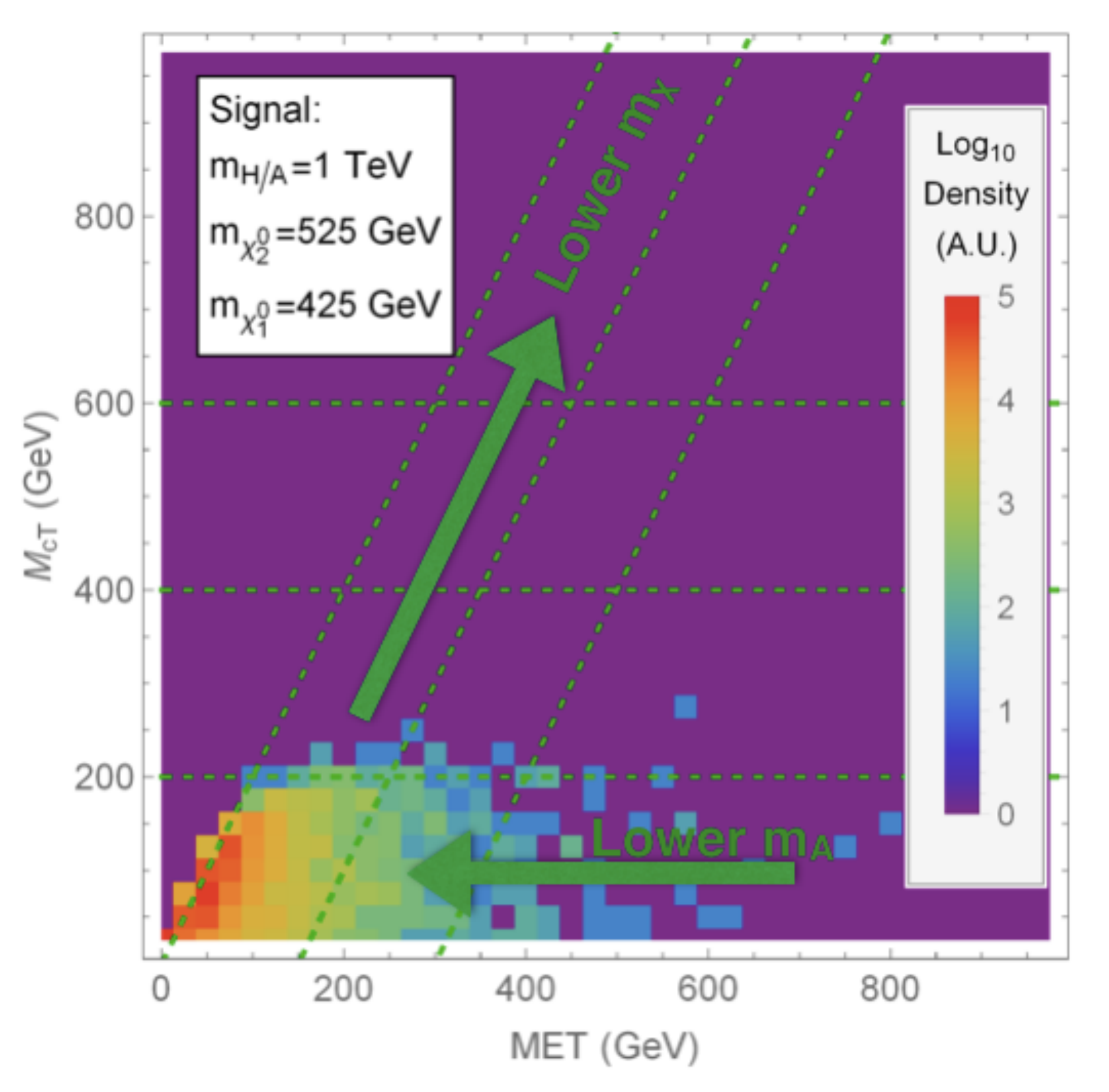}
         }
 \caption{The background (left panel) and signal (right panel) distributions in the $~\met-m_{cT}(\ell\ell,\,\met)$ plane. The signal distribution is shown for a representative benchmark point with $m_H=m_A=1$ TeV, $m_{\chi_1}=425$ GeV, and $m_{\chi_2}=525$ GeV. 
 In the left panel, we show with dashed green lines and arrows the direction of the selection grid described in the text. In the right panel, green arrows indicate how the signal distribution shifts as the heavy Higgs or neutralino masses are varied.}
 \label{Fig:llmet_sigbkg}
 \end{figure*}
 
We make use of two variables to optimize the signal significance: the missing energy $\met$ and the modified clustered transverse mass of the $\ell^+\ell^-+\met$ system, $m_{cT}(\ell\ell,\met)$, defined as
\beq
m_{cT}^2(\ell\ell,\met)=2\times\left((|p_T^{\ell\ell}|+|\metp|)^2-|p_T^{\ell\ell}+\metp|^2\right),
\eeq 
where $\metp$ is the two-vector of the transverse missing energy and $p_T^{\ell\ell}$ is the two-vector sum of the two lepton $p_T$s. In Fig.~\ref{Fig:llmet_sigbkg}, we show the double differential distributions of these two variables for background (left panel)\footnote{In this plot, we include both $ZZ$ and $WW$ backgrounds, even though the $ZZ$ background is dominant.} and signal (right panel) events. 
These distributions suggest that the signal significance can be optimized with a lower cut on $\met$ and an upper cut on $m_{cT}(\ell\ell,\met)$.

For the background distribution, which is dominated by the SM $ZZ$ process, the $\met$ and $m_{cT}(\ell\ell,\met)$ are strongly correlated along the diagonal lines. This can be understood by observing that $p_T^{\ell\ell}$ and $\met$ are equal in size and opposite in direction; in this limit, one obtains $m_{cT}(\ell\ell,\met) \simeq 2\met$ for the background.

This relation, on the other hand, does not hold for the signal events (see Fig.~\ref{Fig:llmet_sigbkg}, right panel). Rather, due to the relatively small mass splitting between the resonant Higgs boson and the neutralino pair (50 GeV for the benchmark point in the plot), the modified clustered transverse mass is significantly reduced, while the $\met$ distribution can spread towards higher values due to hard initial state radiation (that we have obtained through {\tt{PYTHIA}} showering\,\footnote{ In principle, the one-jet matched sample would contain harder initial state radiation and may result in a better signal vs background discrimination.  In this sense, our calculated bounds are conservative.).
}). In the right plot, the green arrows show how the distribution shifts when $m_A$ or $m_\chi\equiv m_{\chi_1}=m_{\chi_2}-100$ GeV are varied. As $m_A$ is reduced,
large $\met$ is harder to achieve. 
As $m_\chi$ is lowered, the system approaches the limit of the SM background where the mass effect becomes increasingly irrelevant, hence $m_{cT}(\ell\ell,\met)$ grows to higher values, approaching the diagonal line reflecting the SM background behavior. For this reason, the bounds get weaker for lower neutralino masses despite higher production cross sections. 

To obtain the LHC reach for this decay process, we sample signal and benchmark points with a 50 (25)~GeV step size for heavy Higgs (neutralino) masses, with a fixed mass splitting $m_{ \chi_{3}}=m_{ \chi_{2}}=m_{\chi_1}+100~\gev$. Following the dilepton invariant mass cut, we construct background and signal differential distributions with 50~\gev~steps along the green dashed lines shown in Fig.~\ref{Fig:llmet_sigbkg} (left panel). 
For each benchmark point generated, we then choose the optimal cuts along these lines to optimize the signal significance with either 300 or 3000 fb$^{-1}$ data. 
This optimization process selects different cuts for signals with different cross sections, with higher cross sections (correlated with lower $\tan\beta$ values) allowing for more aggressive signal selection cuts. The projected sensitivity using this method is shown in Fig.~\ref{Fig:llmet} in the $m_{A}$-$m_{\chi_{2,3}}$ plane for several values of tan\,$\beta$. As anticipated, the reach is broader for lower values of tan\,$\beta$, with HL-exclusion (discovery) limits approaching 900 (700) GeV for tan\,$\beta=5$ (left panel of the figure), and becoming weaker (stronger) for higher (lower) tan\,$\beta$. 

\begin{figure}[t] 
 \vspace{0.cm}
  \center{
    \includegraphics[width=7.5cm]{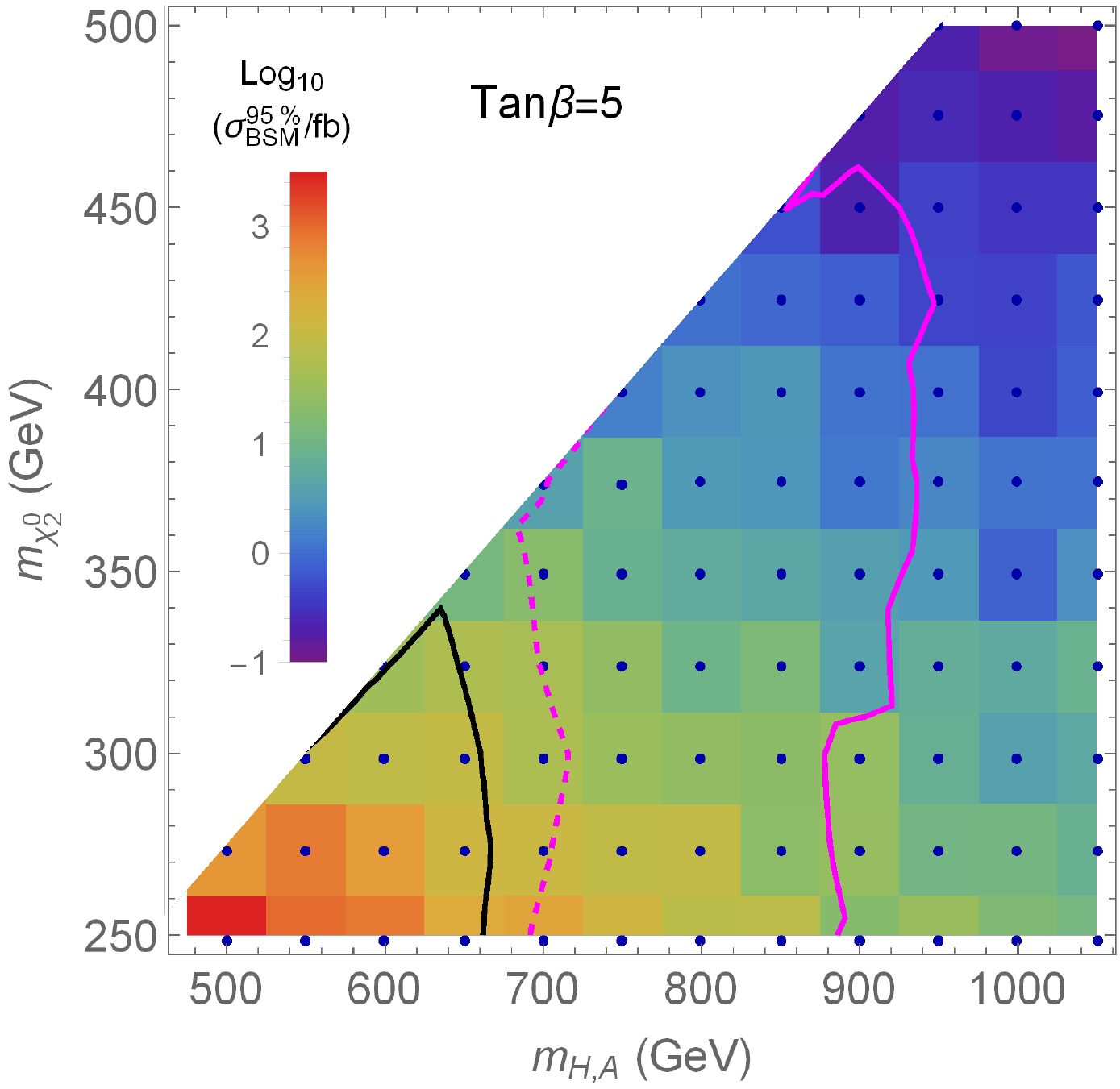}
    \includegraphics[width=7.5cm]{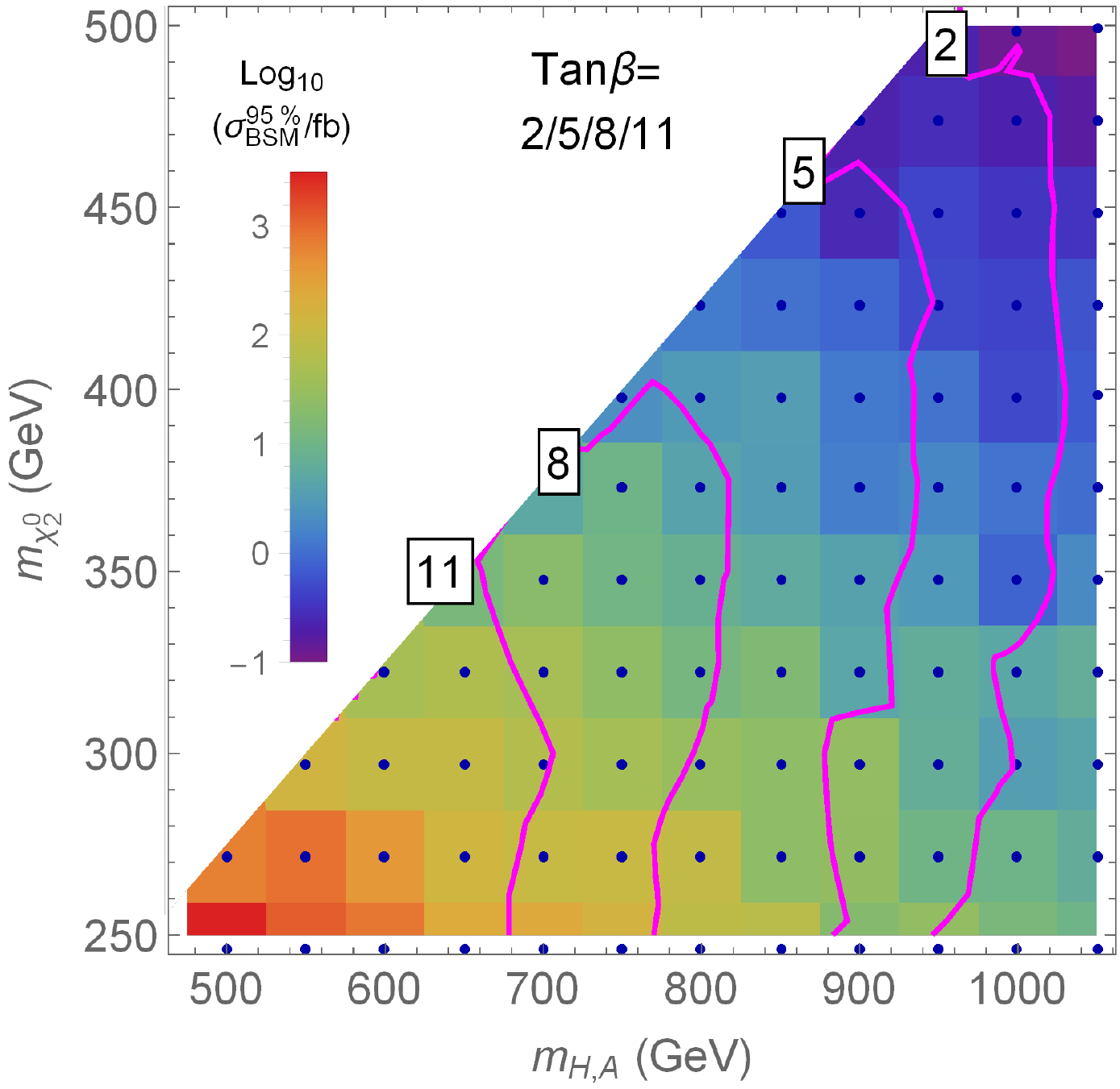}
    }
\caption{{\bf Left Panel}: the projected 2-$\sigma$ exclusion (solid contours) and 5-$\sigma$ discovery lines (dashed contours) at the 13 TeV LHC with integrated luminosities of 300~$\fbi$ (black curves) and 3000~$\fbi$ (magenta curves) for $\tan\beta=5$. {\bf{Right Panel}}: the projected 2-$\sigma$ exclusion with 3000~$\fbi$ data for various $\tan\beta$ values (indicated by the contour labels). For both panels, blue dots indicate simulated signal benchmark points for this analysis, with the colors on each benchmark point tile representing the projected upper limit on the cross section from our analysis for $\tan\beta=5$ as indicated in the plot legend. For the two panels, we fix $m_{\chi_3}=m_{\chi_2}=m_{\chi_1}+100$ GeV.
}
\label{Fig:llmet}
\end{figure}

To compare our reach with existing $Z+\met$ search strategies, we recast the 13 TeV ATLAS $Z+\met$ search from \cite{ATLAS-CONF-2016-056,Aaboud:2017bja} for our signal. While this search is targeted towards events with large $\met$, our signal has relatively lower $\met$ due to the massive LSPs. 
We find that this strategy provides no reach for $\tan\beta\geq3$ even when extrapolated to the HL-LHC, illustrating that existing $Z+\met$ searches, despite targeting the same final states, might not be effective in scenarios with heavy Higgs decays.

In Fig.\,\ref{Fig:chimatbreach}, we show the HL-LHC reach from the proposed search in the $m_{A}$-$\tan\beta$ plane for $m_{\chi_{2,3}}=300$~GeV and $m_{\chi_1}=200$ GeV. Compared to the reach from searches for heavy Higgs decaying into SM fermions, in particular $\tau\tau$ and $t\bar t$ (blue and gray regions in the plot), which are effective at high and low tan\,$\beta$ respectively, \footnote{For some related discussions on heavy Higgs bosons decays into these channels, see Refs.~\cite{Craig:2015jba,Hajer:2015gka,Jung:2015gta,Gori:2016zto,Carena:2016npr,Craig:2016ygr}.} the proposed search is very promising in the intermediate $\tan\beta$ regime where it can probe masses approaching $1$ TeV. This region is out of the reach of these standard searches\footnote{
The reach for Higgsino-like neutralinos from this search can be competitive with direct chargino/neutralino searches (the latter reach can be obtained from extrapolating the current combined direct search results~\cite{CMS-PAS-SUS-17-004}; however, in the compressed regime where the mass splitting is around 100~GeV, the search likely suffers from sizable systematics). }.
Therefore, the proposed search strategy utilizing the exotic decay channels into neutralinos can significantly extend the reach for heavy Higgs bosons to large regions of parameter space not accessible via standard heavy Higgs searches. 
\begin{figure}[t] 
	\vspace{0.cm}
	\center{
		\includegraphics[width=7.5cm]{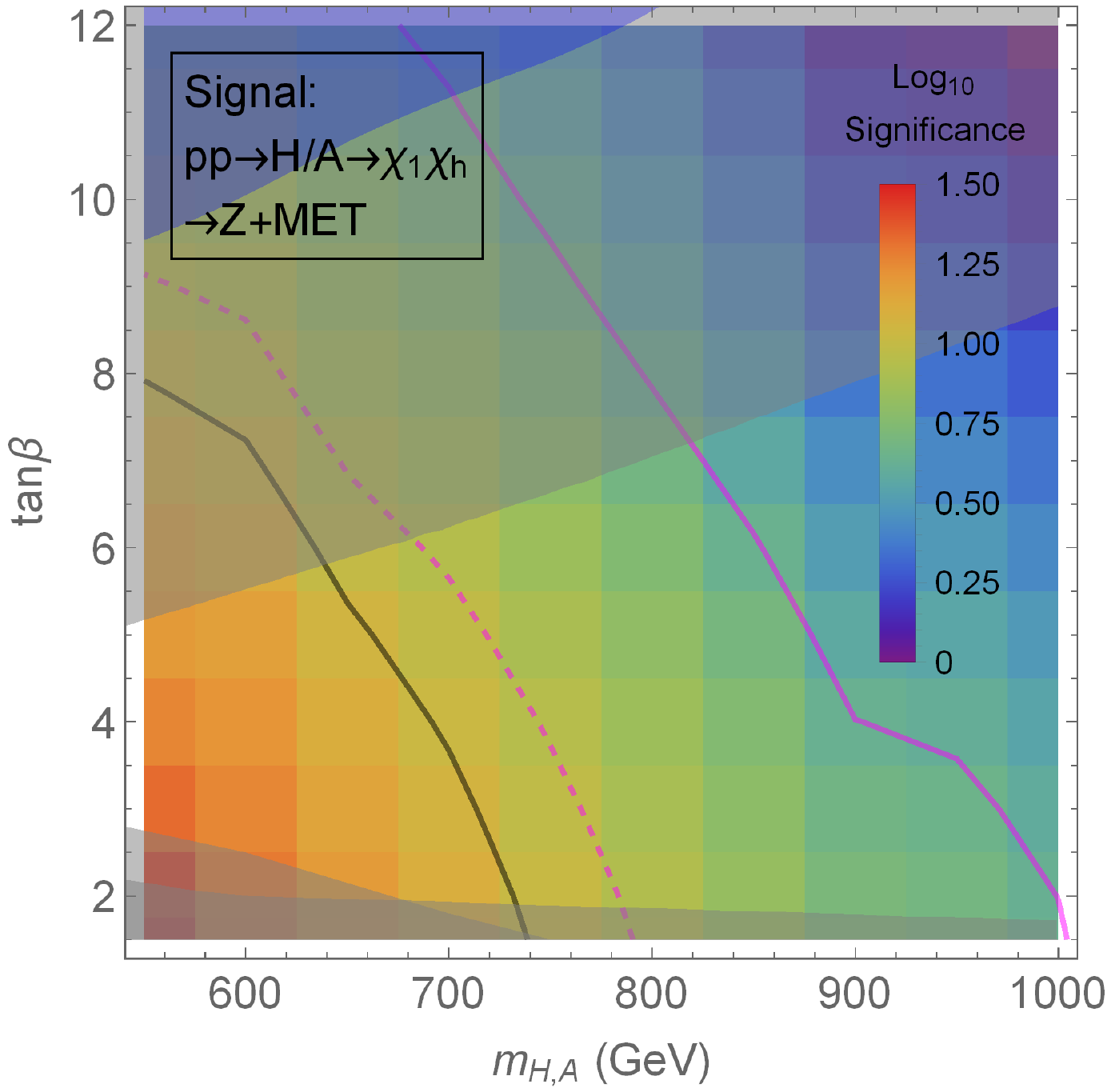}
	}
	\caption{The projected 2-$\sigma$ exclusion limit (magenta) and 5-$\sigma$ discovery reach (dashed magenta) for 13 TeV HL-LHC in the $m_{A}$-$\tan\beta$ plane, for fixed $m_{\chi_{2,3}}=300$~GeV and $m_{\chi_1}=100$ GeV. The solid black curve denotes the corresponding $2\sigma$ exclusion with $300~\fbi$. The colors on each benchmark point tile represent the projected significance of our analysis at 13 TeV HL-LHC. For comparison, we also show the projected limits at HL-LHC for heavy Higgs searches in the $\tau\tau$  ~\cite{CMS-PAS-FTR-16-002} and $t\bar t$~\cite{Carena:2016npr} final states in the upper shaded and lower shaded regions, respectively. The current constraints from $A/H\to\tau\tau$ searches \cite{Aaboud:2017sjh,Sirunyan:2018zut} are shown in the blue shaded region. The lower shaded region that vanishes around 700 GeV represents possible limits from the 4-top final state, for $t\bar t$ associated production of the heavy Higgs ($pp\to t\bar t H/A,~H/A\to t\bar t$) (taken from \cite{Gori:2016zto}).}
	\label{Fig:chimatbreach}
\end{figure}

\section {Searching for heavy Higgs decays to light staus} 
\label{Sec:staus}

In this section, we discuss LHC opportunities to probe light staus in scenarios where they are produced from heavy Higgs decays.\,\footnote{For some earlier studies that also consider heavy Higgs decays to staus, see \cite{Medina:2017bke,Arganda:2018hdn}.} As mentioned previously, direct LHC bounds on staus \cite{Sirunyan:2018vig} are still weaker than the corresponding LEP bounds \cite{Abdallah:2003xe} due to relatively small production cross sections and large backgrounds. Hence heavy Higgs boson decays could offer exciting opportunities to probe staus at the LHC. 

\subsection{Benchmark models}

As discussed in Sec. \ref{sec:2.3}, heavy Higgs bosons can have significant branching fractions to light staus for large values of $A_{\tau} t_\beta/v$ (see couplings in Eq.\,\ref{eq:staucouplings}). This coupling grows at large tan\,$\beta$ as expected for a down-type Yukawa coupling, leading to significantly higher stau production cross sections compared to direct production~(see the colors in the right panel of Fig.\,\ref{Fig:Branchingratios}). 

\begin{figure*}[t] 
 \vspace{0.cm}
  \center{
\includegraphics[width=10cm]{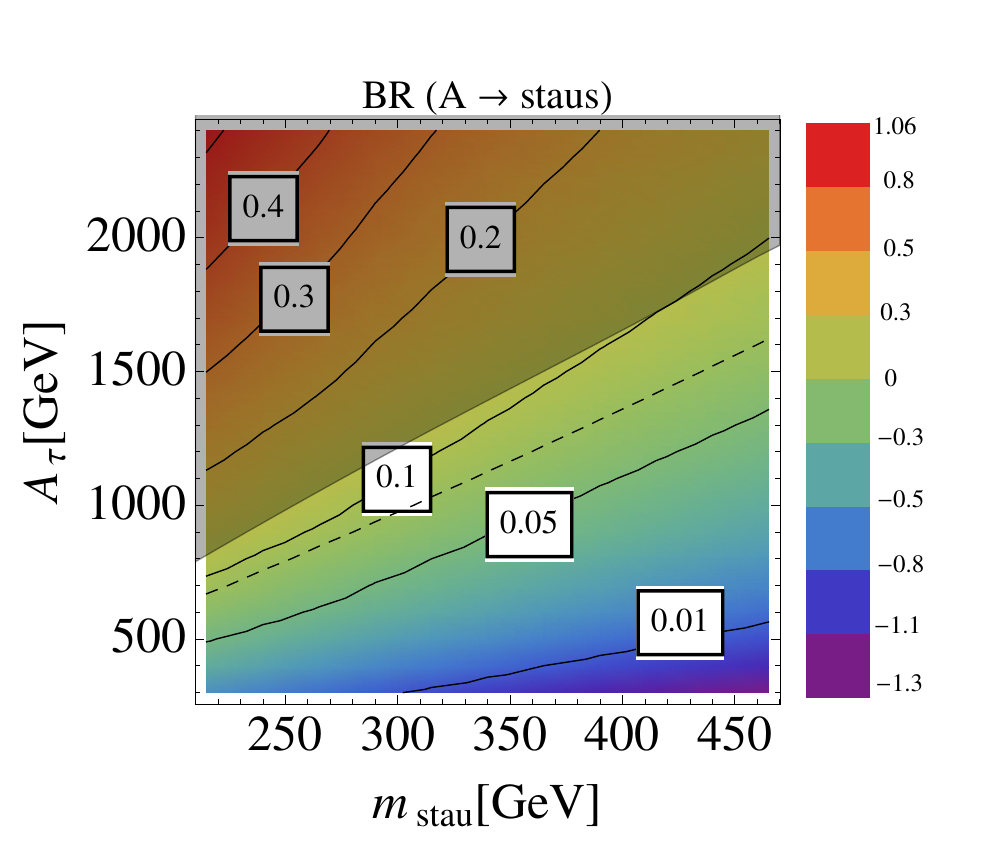}}
\caption{Solid contours with numerical labels represent the branching ratio BR$(A\to$ staus), for tan\,$\beta=10$, $\mu=500$ GeV, $m_{\tilde{\tau}_L}=m_{\tilde{\tau}_R}=m_{\text{stau}}$,\,$m_A=2\,m_{\text{stau}}+70$ GeV, $M_1=100$ GeV, and all other dimensionful parameters set to 2 TeV. The color coding denotes the ratio of branching ratios into staus to those into taus, log$_{10}\left(\frac{BR(A\to \tilde{\tau}\tilde{\tau})}{BR(A\to\tau\tau)}\right)$, and the dashed curve indicates where these two are equal. The shaded region is  constrained by the requirement of absolute stability of the electroweak vacuum. The heavy scalar, $H$, leads to similar results for the branching ratios.}
\label{Fig:stauBR}
\end{figure*}

To illustrate the dependence on $A_{\tau}$, in Fig.\,\ref{Fig:stauBR} we plot the branching ratio BR$(A\to$ staus)\,\footnote{For $A_{\tau} t_\beta \gg \mu$, as is the case in this plot, the stau couplings to both $A$ and $H$ become equal (see Eq.\,\ref{eq:staucouplings}), resulting in similar values for BR$(H\to$ staus).} as a function of the stau soft masses, $m_{\tilde{\tau}_L}=m_{\tilde{\tau}_R}=m_{\text{stau}}$, and $A_{\tau}$, for fixed $t_\beta=10$, $\mu=500$ GeV, and $m_A=2\,m_{\text{stau}}+70$ GeV (this results in $m_A\approx2\,m_{{\tau}_2}+50$ GeV, used in our collider analysis presented in the next section), $M_1=100$ GeV, and all other dimensionful parameters set to 2 TeV. Note that for $m_{\tilde{\tau}_L}^2=m_{\tilde{\tau}_R}^2\,>\,m_{\tau}|A_{\tau}-\mu\tan\beta|$, the stau mass eigenstates are approximately degenerate, and the distinction between $\tilde{\tau}_1$ and $\tilde{\tau}_2$ becomes irrelevant for collider purposes. 
Fig.\,\ref{Fig:stauBR} clearly illustrates that BR$(A\to$ staus) can be steadily increased by increasing the value of $A_{\tau}$, with branching ratios approaching $\sim50\%$ for $A_{\tau}\sim 2500 $~GeV. The color coding in the plot denotes the ratio of the branching ratio of $A$ to staus to the branching ratio into taus, log$_{10}([BR(A\to \tilde{\tau}\tilde{\tau})]/[BR(A\to\tau\tau)])$: we see that the branching ratios to staus can be much larger at large $A_{\tau}$, in some cases by more than an order of magnitude.

For very large values of $A_\tau$, however, the stability of the electroweak vacuum becomes an important constraint, since new charge breaking minima where the stau fields acquire a vacuum expectation value can appear \cite{Rattazzi:1996fb,Hisano:2010re,Sato:2012bf,Kitahara:2012pb,Carena:2012mw,Medina:2017bke}. In our study, we require absolute stability of the electroweak vacuum by numerically solving for the minima of the Higgs-stau scalar potential and verifying that the electroweak vacuum is the deepest minimum; particularly, we take the tree level scalar potential of the four fields $H_u, H_d, \tilde\tau_L,\tilde \tau_R$, and include the leading 1-loop contribution from top/stop loops as discussed in Ref.\,\cite{Carena:2012mw}. The requirement of absolute vacuum stability excludes the shaded region in Fig.\,\ref{Fig:stauBR}, constraining BR$(A/H\to$ staus) to $\lesssim 12\%$ and $\frac{BR(A\to \tilde{\tau}\tilde{\tau})}{BR(A\to\tau\tau)} \lesssim 1.4$.\,\footnote{In principle, the electroweak vacuum need not be absolutely stable. One could relax the bound on $A_\tau$ by imposing the vacuum to be metastable with a lifetime longer than the age of the universe. We estimate that this could allow up to $\sim 50\%$ larger values of $A_\tau$, and therefore $BR(A\to \tilde{\tau}\tilde{\tau})$ as large as $\sim 20\%$.}

At this point, it is illuminating to observe the parallels between $A/H\to$ staus and $A/H\to\tau\tau$. Both branching ratios increase at larger values of tan\,$\beta$, where absolute vacuum stability forces BR$(A/H\to$ staus)$\,\sim1.4\times\,$BR$(A/H\to\tau\tau)$ at best. If staus are degenerate and are the next-to-lightest supersymmetric particles, they decay as $\tilde{\tau}_{1,2}\to \tau\chi_1$, so that $A/H\to$ staus and $A/H\to\tau\tau$ give the same visible final states at the LHC (but with different amounts of missing energy). The $A/H\to\tau\tau$ channel, in particular for b-associated production at large tan$\,\beta$, currently provides the strongest bounds on heavy Higgs bosons in the MSSM \cite{Aaboud:2017sjh,Sirunyan:2018zut}. The above similarities between the two decay channels suggest that heavy Higgs produced in association with b-quarks and decaying into staus can likewise be promising avenues with similar search strategies. This setup offers two interesting directions: 
\begin{itemize}
\item $m_{\tilde{\tau}}\ll m_A/2$: The staus are highly boosted, giving similar kinematic distributions to $\tau$'s and little missing energy if the LSP is effectively massless (as the two LSPs produced are back-to-back in the lab frame). In this scenario, the $A/H\to$ staus signature will be efficiently captured by the LHC searches for $A/H\to\tau\tau$, hence large BR$(A/H\to$ staus) will 
broaden the reach for heavy Higgs bosons in the $(m_A-$tan\,$\beta)$ plane towards lower values of $\tan\beta$ with the same search strategy. 

More interesting is the case of a small mass gap between the $\tilde \tau$'s and the LSP, where standard searches of staus from direct production lose sensitivity since the stau decay products are soft. In this scenario, the boost provided from the heavy Higgs decay can result in the $\tau$ leptons produced from stau decay passing the detection threshold. Additional kinematic handles~\cite{Lester:1999tx,Han:2012nm,Han:2012nr,Fox:2012ee,Agashe:2013eba,An:2015uwa,Macaluso:2015wja}, or reconstruction of the heavy resonance, 
could then enhance the reach for such topologies. We leave this study for future work.
\item Heavier $m_{\tilde{\tau}}$: The $\tau$ kinematical distributions will be significantly softer than the corresponding distributions from $H\to\tau\tau$ decay, as shown in the left panel of Fig.~\ref{Fig:taupt}. In addition, such scenarios also involve sizable missing energy and large $m_{T2}$ from the intermediate stau states, as shown in the right panel of Fig.~\ref{Fig:taupt}. Furthermore, as we will discuss next, with this mass spectrum, stau production via heavy Higgs decays can have significantly larger cross sections than stau direct production, thus serving as the dominant source of staus at the LHC. 
\end{itemize}
In this paper, we will focus on the second case, as this is an obvious target for novel LHC search strategies, as well as offering a significant enhancement on the reach for staus. 

\begin{figure*}[t] 
 \vspace{0.cm}
  \center{\includegraphics[width=7.5cm,height=5cm]{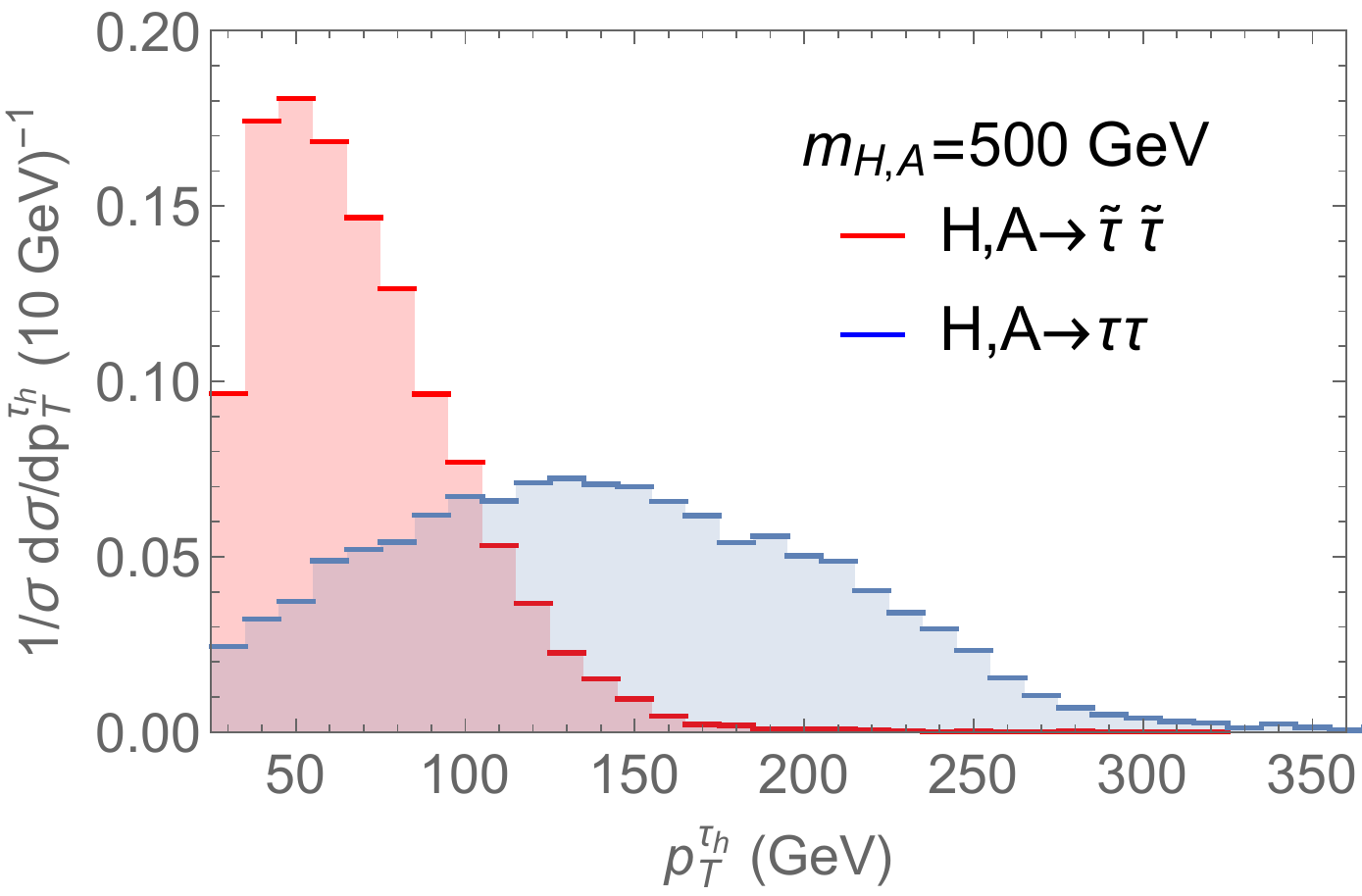}~~~
  \includegraphics[width=7.9cm, height=4.9cm]{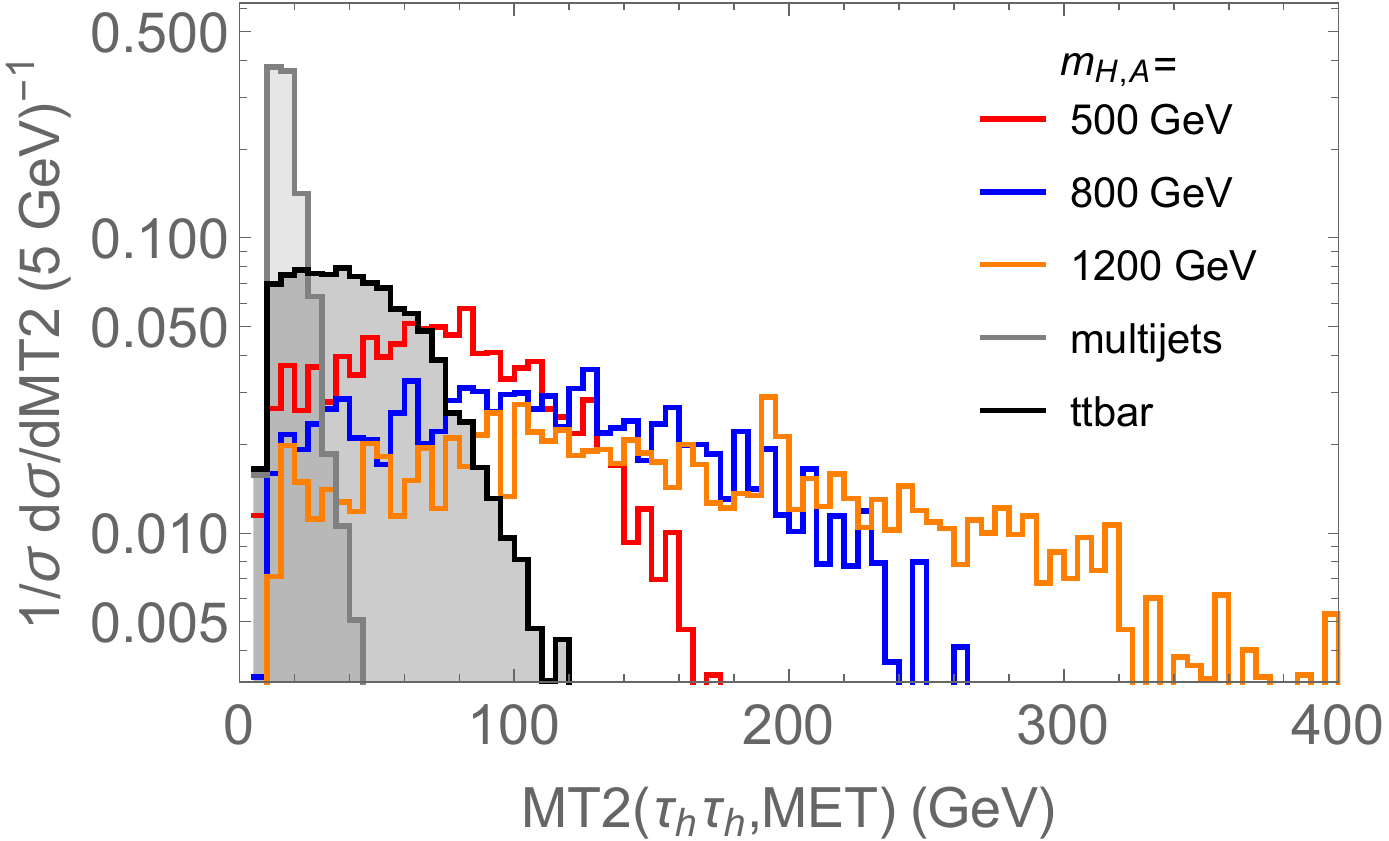}}
  \caption{{\bf{Left Panel}}: The differential $p_T$ distribution of the hadronic $\tau$ jets produced directly from $H,A\to \tau\tau$ (blue) and from stau decays from $H,A\to \tilde \tau \tilde \tau$ (red), in the $b\bar b$ associated production of the heavy Higgs bosons, with $m_{H,A}=500$ GeV, $m_{\tilde{\tau}}=225$ GeV, and $m_{\chi_1}=100$ GeV. {\bf{Right Panel:}} The differential $m_{T2}$ distribution of the hadronic $\tau$ jets from $H,A\to \tilde \tau \tilde \tau$ from $b\bar b$ associated production. We show the distribution for $m_{H,A}=500,800,1200$ GeV, $m_{\tilde{\tau}}=225,375,575$ GeV in red, blue, orange, respectively, with $m_{\chi_1}=100$ GeV. The gray shaded regions show the corresponding differential distributions for the two leading backgrounds from SM multijet process (lighter gray) and $t\bar t$ (darker gray). }
\label{Fig:taupt}
\end{figure*}

We consider the mass spectrum $m_A= 2\,m_{\tilde{\tau}_2}+50$ GeV and $M_1=100$ GeV. We take $\mu=\text{max}(500,\,m_{\tilde{\tau}}+25)$ GeV so that the Higgsino-like neutralinos and charginos remain heavier than the staus\,\footnote{There also exist LHC constraints on charginos and neutralinos decaying to staus \cite{CMS-PAS-SUS-17-002}. The parameter space we consider in this paper is compatible with such bounds.} (and also remain out of reach of the LHC), thus the staus decay as $\tilde{\tau}_{1,2}\to \tau\chi_1$ $100\%$ of the time. We decouple the remaining SUSY particles from the spectrum and fix the value of $A_{\tau}$ to the largest value compatible with the requirement of absolute vacuum stability as discussed above. The combined production cross section of staus via decays of both $A$ and $H$, where these are produced in association with b-quarks, $\sigma(pp\to b\bar b(A,H\to$\,staus$))$ is plotted in the $m_A-$tan$\beta$ plane in Fig.\,\ref{Fig:stauCS}. The gray shaded region at large $\tan\beta$ is the region already probed by LHC $A/H\to\tau\tau$ searches. In the region allowed by current constraints, cross sections up to $\sim50$ fb are possible for $m_A=500$ GeV. In the plot, the color coding represents the ratio of production cross section from heavy Higgs decay to direct production cross section of maximally mixed staus,
$\sigma(pp\to b \bar b(A,H\to$\,staus$))/\sigma(pp\to$\,staus).  Along the dashed black line, these two cross sections are equal. The production from heavy Higgs decay can be the dominant source of staus for tan\,$\beta\gtrsim5$, and can be more than an order of magnitude larger in the allowed region of parameter space, thus offering a promising avenue to probe light staus beyond what is accessible via direct production.

\subsection {Proposed search}

We now discuss a new search strategy for staus produced from heavy Higgs boson decays. We begin with the current ATLAS search \cite{Aaboud:2017sjh} for heavy Higgs bosons decaying into tau leptons,\,\footnote{The corresponding CMS search is available in Ref.\,\cite{Sirunyan:2018zut}.} and modify the strategy in order to make use of the additional kinematic handles available in the decay to staus.  

\begin{figure*}[t] 
 \vspace{0.cm}
  \center{
  \includegraphics[width=10cm]{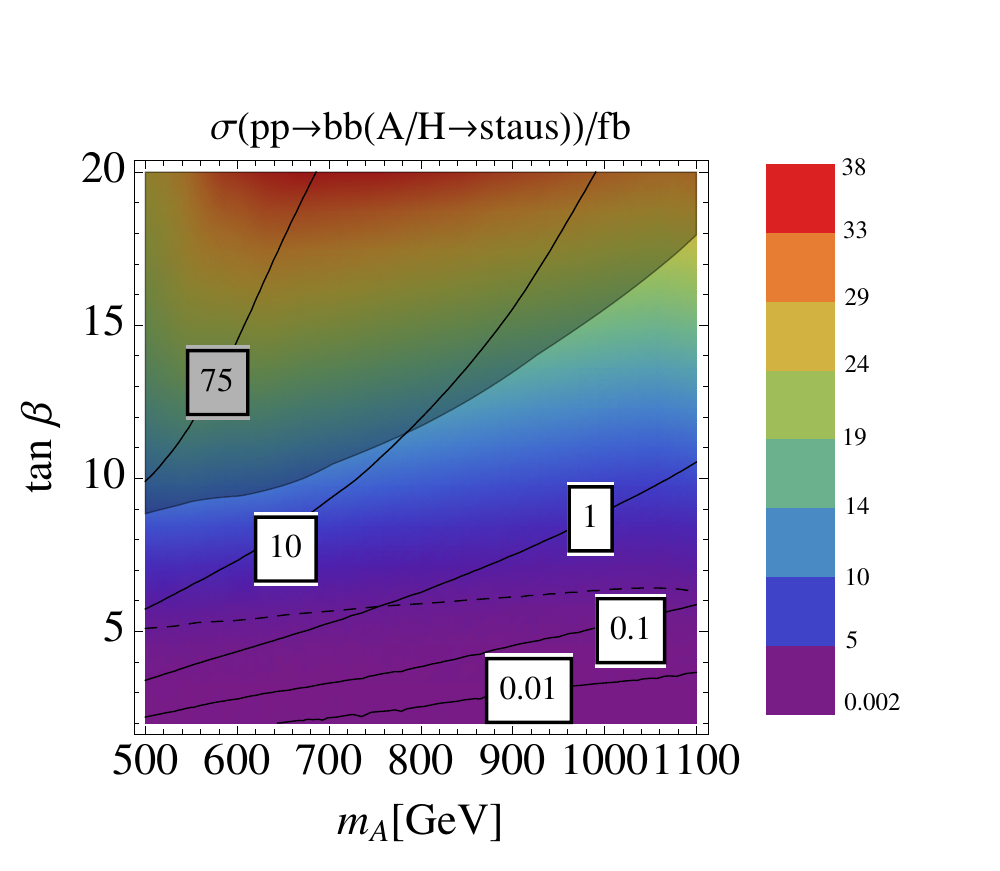}
  }
\caption{Solid contour lines with numerical labels denote the production cross section of staus from b-associated production and decay of $A,H$ (combined) in fb. For this plot we fix $m_A= 2m_{\tilde{\tau}_2}+50$ GeV, $\mu=\text{max}(500,\,m_{\tilde{\tau}}+25)$ GeV, and set $A_{\tau}$ to the maximum value allowed by the absolute vacuum stability bound at each point. The color coding denotes the ratio of production cross section of staus from this decay process to the direct production cross section of maximally mixed staus, $\sigma(pp\to b \bar b(A,H\to$\,staus$))/\sigma(pp\to$\,staus). The dashed curve denotes the contour along which these two production cross sections are equal. The shaded region is ruled out from the current $A/H\to \tau \tau$ LHC searches \cite{Aaboud:2017sjh,Sirunyan:2018zut}.}
\label{Fig:stauCS}
\end{figure*}

The ATLAS search \cite{Aaboud:2017sjh} is very inclusive and does not require any tau-pair resonance reconstruction. The search has four signal categories, focusing on leptonic taus and hadronic taus, with and without b-tagged jet for $b\bar b$ associated Higgs production and gluon fusion, respectively. (1) For the leptonic tau signal regions, the search requires two opposite-sign leptons with $p_T > 30$ GeV in the central region $|\eta|<2.4$, vetoing additional leptons and the invariant mass of this pair of leptons in the 80-110~\gev\ range to remove the $Z$-boson background. In addition, the transverse mass of an individual lepton plus the $\met$ system is required to be smaller than 40~\gev\ to remove $W$ boson related backgrounds. For our signal, the tau leptons from stau decays are softer, making it more difficult to satisfy the hardness of the lepton requirement. Furthermore, the decays from the staus have sizable transverse mass, failing the low transverse mass requirement. (2) For the hadronic tau searches, the analysis makes use of the resonance nature of the tau lepton pairs, imposing hard requirements on their $p_T$ (leading hadronic tau lepton $p_T>130$~GeV and subleading tau lepton $p_T>60$ GeV), and further requires them to be back-to-back ($\Delta\phi > 2.4$). These requirements cannot be easily satisfied by our signal events (see the left panel of Fig. \ref{Fig:taupt}). Therefore, the ATLAS search \cite{Aaboud:2017sjh} is not suited for probing staus from heavy Higgs decays. Nevertheless, recasting this analysis and understanding the backgrounds offers significant insight into this search. Building on this, we develop a new search strategy for the $H,A\to \tilde \tau \tilde \tau$ signal that utilizes the missing energy and sizable $m_{T2}$ variable present in the signal events. We recast four different search categories: both hadronic and leptonic $\tau$, with and without an additional b-tagger, successfully reproducing both the signal and background yield presented by the ATLAS collaboration~\cite{Aaboud:2017sjh}. For moderate to large $\tan\beta$, we find the hadronic $\tau$ with additional $b$-jet tag to be the most sensitive channel, hence we only show the numerical results only for this signal region.

The cut-flow information from this recast and our devised cuts on our signal topology $H\to \tilde \tau \tilde \tau$ as well as the major backgrounds are presented in Table.~\ref{tab:stau}. For the $\tilde{\tau}$ decay signals, we use two benchmark values of the heavy Higgs bosons masses, 500 and 1000 GeV. For the stau signals, we additionally use ($m_A=500$ GeV, $m_{\tilde{\tau}}=225$ GeV) and ($m_A=1000$ GeV, $m_{\tilde{\tau}}=475$ GeV), which have BR$(A\to$staus)$\approx$ 15\% and 9\%, respectively (and similar branching ratios for H), for $\tan\beta=10$ and $A_\tau$ fixed to the vacuum stability bound. The LSP mass is set to 100 GeV. 

\begin{table}[t]
	\centering
	\begin{tabular}{|c|c|c|c|c|c|c|c|c|c|}
	\hline
		\multicolumn{2}{|c|}{}	& \multicolumn{4}{|c|}{Projection of the $H\rightarrow \tau\tau$ search \cite{Aaboud:2017sjh}} & \multicolumn{4}{|c|}{New Search} \\ \cline{3-10}
		\multicolumn{2}{|c|}{}	& \multicolumn{2}{|c|}{Background} & \multicolumn{2}{|c|}{Signal $m_A$/GeV}  & \multicolumn{2}{|c|}{Background} & \multicolumn{2}{|c|}{Signal $m_A$/GeV} \\ \hline
		\multicolumn{2}{|c|}{Selection cuts}	& \multicolumn{1}{|c|}{multijet} & \multicolumn{1}{|c|}{$t\bar t$} & \multicolumn{1}{|c|}{500} & \multicolumn{1}{|c|}{1000}  & \multicolumn{1}{|c|}{multijet} & \multicolumn{1}{|c|}{$t\bar t$} & \multicolumn{1}{|c|}{500} & \multicolumn{1}{|c|}{1000} \\ \hline
		\multicolumn{2}{|c|}{Baseline} & \multicolumn{1}{|c|}{1.7$\cdot10^9$} & \multicolumn{1}{|c|}{1.3$\cdot$10$^7$} &    2.3$\cdot$10$^5$   &   4.6$\cdot$10$^3$  & \multicolumn{1}{|c|}{1.7$\cdot10^9$} & \multicolumn{1}{|c|}{1.3$\cdot10^7$} & \multicolumn{1}{|c|}{2.3$\cdot10^5$} & \multicolumn{1}{|c|}{4.6$\cdot10^3$} \\ \hline
		\multicolumn{2}{|c|}{2 tagged $\tau_h$} & \multicolumn{1}{c}{2.6$\cdot10^7$} & \multicolumn{1}{|c|}{8.8$\cdot10^5$} &  2.7$\cdot$10$^4$   &   650 & \multicolumn{1}{|c|}{2.6$\cdot10^7$} & \multicolumn{1}{|c|}{8.8$\cdot10^5$} & \multicolumn{1}{|c|}{2.7$\cdot10^4$} & \multicolumn{1}{|c|}{650} \\ \hline
		\multicolumn{2}{|c|}{$p_T^{\tau_{1,2}}\!>\!130,65\,$GeV} & \multicolumn{1}{|c|}{9.0$\cdot10^4$} & \multicolumn{1}{|c|}{2.7$\cdot10^4$} &   720    &  330     & \multicolumn{1}{|c|}{---} & \multicolumn{1}{|c|}{---} & \multicolumn{1}{|c|}{---} & \multicolumn{1}{|c|}{---} \\
		\multicolumn{2}{|c|}{$p_T^{\tau_{1,2}}>35~\gev$} & \multicolumn{1}{|c|}{---} & \multicolumn{1}{|c|}{---} &   ---    &   ---    & \multicolumn{1}{|c|}{4.9$\cdot10^6$} & \multicolumn{1}{|c|}{3.4$\cdot10^5$} & \multicolumn{1}{|c|}{1.2$\cdot10^4$} & \multicolumn{1}{|c|}{380} \\ \hline
		\multicolumn{2}{|c|}{OS \& $\ell$-veto} & \multicolumn{1}{c}{1.6$\cdot10^4$} & \multicolumn{1}{|c|}{1.4$\cdot10^4$} &   420    &  270    & \multicolumn{1}{|c|}{1.0$\cdot10^6$} & \multicolumn{1}{|c|}{3.3$\cdot10^5$} & \multicolumn{1}{|c|}{9.4$\cdot10^3$} & \multicolumn{1}{|c|}{300} \\ \hline
		\multicolumn{2}{|c|}{$\Delta(\phi^{\tau_1}, \phi^{\tau_2})>2.4$} & \multicolumn{1}{l}{9.2$\cdot10^3$} & \multicolumn{1}{|c|}{8.6$\cdot10^3$} &   140    &  55   & \multicolumn{1}{|c|}{---} & \multicolumn{1}{|c|}{---} & \multicolumn{1}{|c|}{---} & \multicolumn{1}{|c|}{---} \\
		\multicolumn{2}{|c|}{$\Delta(\phi^{\tau_1}, \phi^{\tau_2})>0.4$} & \multicolumn{1}{|c|}{---} & \multicolumn{1}{|c|}{---} &   ---    &  ---   & \multicolumn{1}{|c|}{1.0$\cdot10^6$} & \multicolumn{1}{|c|}{2.2$\cdot10^5$} & \multicolumn{1}{|c|}{9.4$\cdot10^3$} & \multicolumn{1}{|c|}{300} \\ \hline
		\multicolumn{2}{|c|}{b-tagged jet} & \multicolumn{1}{l}{6.9$\cdot10^3$} & \multicolumn{1}{|c|}{6.2$\cdot10^3$} & 70 & 34 & \multicolumn{1}{|c|}{1.8$\cdot10^5$} & \multicolumn{1}{|c|}{1.7$\cdot10^5$} & \multicolumn{1}{|c|}{4.4$\cdot10^3$} & \multicolumn{1}{|c|}{190} \\ \hline
 \multicolumn{2}{|c|}{MET $>$ 200 GeV} &   ---    &   ---    &   ---    & --- &9.2$\cdot 10^4$ & 1.2$\cdot10^5$ & 3.4$\cdot10^3$ & 170 \\ \hline
\multirow{3}[0]{*}{MT2} & $>$\,40\,GeV &  ---  &	--- &	--- &  ---  & 7.7$\cdot10^3$ & 5.5$\cdot10^4$ & 2.3$\cdot10^3$ & 140 \\ \cline{2-10}
  & $>$\,80\,GeV &   ---    &  ---  & ---	&--- & 480 & 7.7$\cdot10^3$ & 1.1$\cdot10^3$ & 105 \\ \cline{2-10}
  & $>$\,120\,GeV &    ---   &  --- &---	&---	&  95 & 570 & 230 & 74 \\ \hline
	\end{tabular}%
	\caption{Cut-flow table for the signal and major backgrounds for the LHC $H\to \tau\tau$ search (hadronic, b-tag signal region) \cite{Aaboud:2017sjh} and our proposed search for $H\to \tilde \tau \tilde \tau$ in the $b\bar b$ associated production channel of the heavy Higgs bosons. The entries represent the expected number of events at the HL-LHC. For the signal yield, we choose $\tan\beta=10$. The two stau benchmark points correspond to ($m_A=500$ GeV, $m_{\tilde{\tau}}=225$ GeV) and ($m_A=1000$ GeV, $m_{\tilde{\tau}}=475$ GeV), and have BR$(A\to$staus)$\approx$ 15\% and 9\% respectively (similar for H), while $M_1=100$ GeV. }
	\label{tab:stau}%
\end{table}%

Let us first discuss the ATLAS analysis (first columns in the table, ``Projection of the $H\to\tau\tau$ search \cite{Aaboud:2017sjh}''). The major backgrounds are from SM multijet processes and top quark pair production. After the generation of the signal and background events with basic selection cuts for collider acceptance (baseline entries in the table), the signal to background ratio is $\sim10^{-6}\,(10^{-4})$,  for $m_A=1000\,(500)$ GeV, with the background dominated by multijet process. Requiring two tagged hadronic tau leptons with $p_T$ cuts improves the ratio by three orders of magnitude. However, other selection cuts such as opposite-sign hadronic tau pairs, additional lepton vetos, and an additional $b$-tagged jet do not further improve this ratio significantly
\footnote{These cuts are helpful in reducing backgrounds from other processes, such as multi-boson productions, not discussed here.}.

As mentioned earlier, the crucial differences between the stau signal of interest and the $H\to\tau\tau$ signal are that the $\tau$ leptons are softer and not back-to-back in the lab frame. Consequently, we need to relax the $\tau$ $p_T$ requirement (changing the cut from 130, 65 GeV to 35 GeV),
as well as the requirement that the tau leptons be back-to-back (changing the $\Delta\phi$ cut from 2.4 to 0.4), in order to retain the majority of the signal events. However, these relaxed cuts result in a larger background ($\sim 30$ times larger than the background in the ATLAS $H\to\tau\tau$ search after the stronger $\tau$ $p_T$ and $\Delta\phi$ cuts). To reduce this background, we additionally make use of the $\met$ and $m_{T2}$ variables. To this effect, we impose a
minimal $\met$ cut of 200 GeV, and use three cuts on $m_{T2}>\!40, 80,$ and $120$ GeV, aiming at low, intermediate and heavy Higgs masses, respectively. 
The table shows that these cuts are very efficient at reducing background while preserving the signal events. 

\begin{figure*}[t] 
	\vspace{0.cm}
	\center{
		\includegraphics[width=10cm]{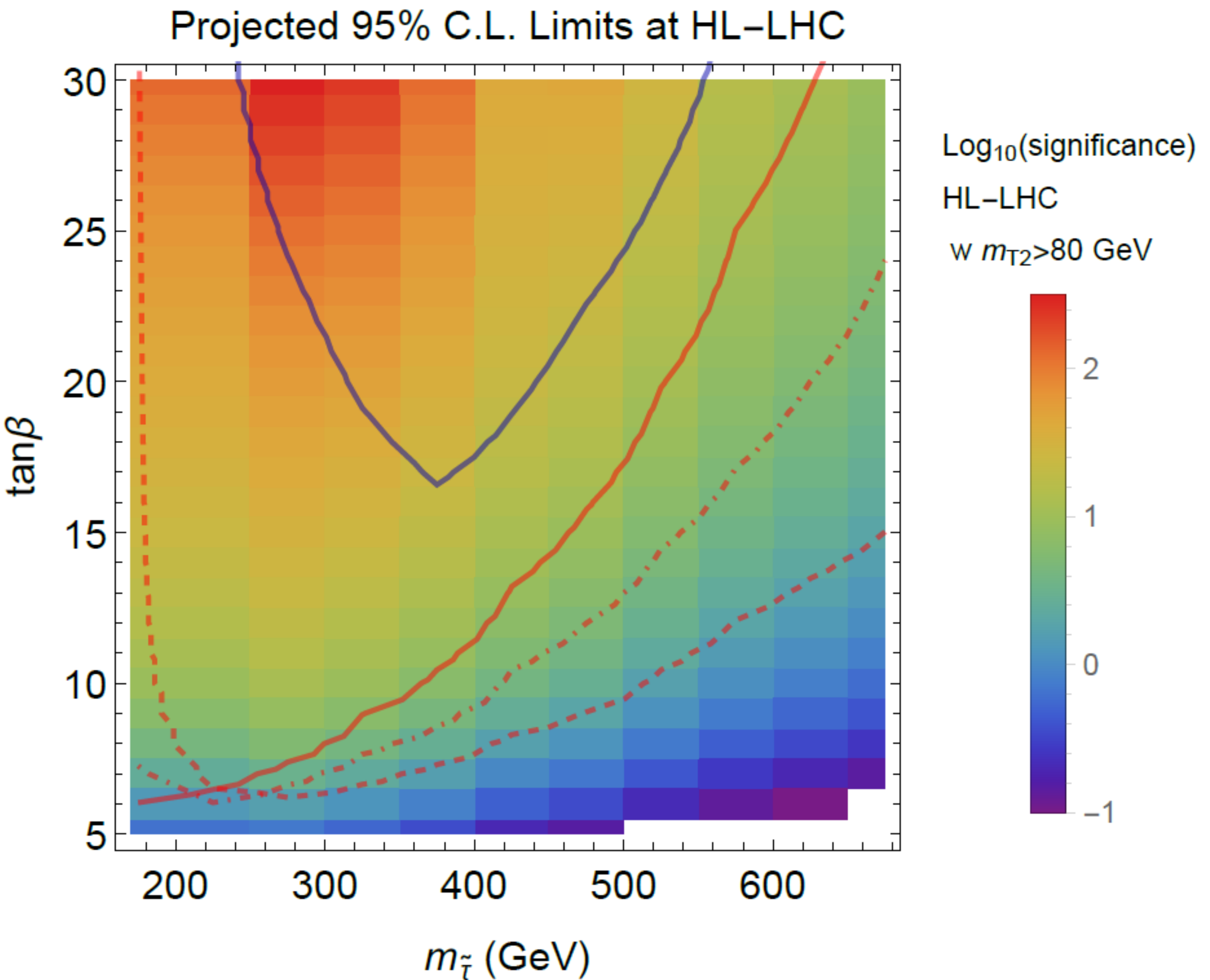}
	}
	\caption{Projected 95\% C.L. exclusion sensitivity at the HL-LHC, for $m_A=2m_{\tilde \tau_2}+50$ GeV  and $m_{\text{LSP}}=100$ GeV, using the cuts outlined in Table\,\ref{tab:stau}. The solid, dot-dashed, and dotted red curves correspond to the proposed search with $m_{T2}>\!40, 80,120$ GeV, respectively. The blue curve corresponds to the sensitivity derived from applying the heavy Higgs to $\tau\tau$ search strategy~\cite{Aaboud:2017sjh} to our Higgs decay to stau signal sample, projected to HL-LHC with $3~\abi$ of integrated luminosity. The colors show the significance using the $m_{T2}> 80$ GeV signal region. 	}
	\label{Fig:staulimits}
\end{figure*}

The expected 95\% C.L. limit from this search strategy for $m_A=2m_{\tilde \tau_2}+50$ GeV, $m_{\text{LSP}}=100$ GeV is shown in Fig.~\ref{Fig:staulimits} in the $m_{\tilde \tau}$-$\tan\beta$ plane for the various $m_{T2}$ cuts as outlined in Table\,\ref{tab:stau}. The softer $m_{T2}> 40$ GeV cut (solid red line) preserves more signal events in the low $m_A$ region, dominating the limits in the $m_{\tilde \tau}\sim180-250$~GeV range. In contrast, the harder $m_{T2}$$> 120$ GeV
cut (dashed red line) can significantly suppress the background, thus drives the limits in the large $m_{\tilde \tau}$ range. For comparison, in the figure the blue solid line shows the projected HL-LHC limits from the  recast and projection of the ATLAS $A,H\to \tau\tau$ search \cite{Aaboud:2017sjh} on our $A,H\rightarrow \tilde \tau \tilde \tau$ signal sample, showing that the new proposed searches improve the coverage significantly. Furthermore, using the information on signal significance encoded in the color coding in the figure together with the dependence of the relevant couplings on $A_{\tau}$ (see Eq.\,\ref{eq:staucouplings}), one can estimate how the reach changes as $A_{\tau}$ is varied away from the vacuum stability limit. 

Fig.~\ref{Fig:staulimits} shows that the above search strategy can probe light staus in a large region of parameter space not accessible via direct searches. Furthermore, this search strategy can also contribute to the reach for heavy Higgs bosons in the $(m_A-\tan\beta)$ plane for these stau benchmark scenarios. We find that the $m_A$ reach from the analysis with the strongest cut, $m_{T2}>120$ GeV (dotted red curve in this figure) is comparable to the projected reach from the standard b-associated $H\to\tau\tau$ search \cite{CMS-PAS-FTR-16-002}, suggesting that $A,H\to$ staus can be an important discovery/co-discovery channel for the heavy Higgs bosons in addition to $A,H\to\tau\tau$.\,\footnote{Recall that for these studies we set $A_\tau$ to the maximum values allowed by vacuum stability; for metastable vacua with lifetime longer than the age of the Universe, $A_\tau$ and consequently BR($A/H\to$ staus) can be larger, and $A/H\to$ staus can potentially even dominate over $A,H\to\tau\tau$ as the leading discovery channel.}

\section{Other decay channels and signatures}
\label{Sec:others}

In this section, we offer a brief overview of additional heavy Higgs decay signals arising in the MSSM that could be viable at the HL-LHC. 
This is not meant to be an exhaustive list, and we only discuss these possibilities at a qualitative level, without performing collider analyses of the possible reach. Moreover, a wider variety of signals are possible in other non-minimal scenarios, such as away from the alignment/decoupling limit, with cascade decays in the neutralino sector (e.g. $A/H\to \chi_2 \chi_3$, with both of these neutralinos then decaying to the LSP), or with additional particles available (such as the wino). A more comprehensive discussion of these is beyond the scope of this paper.

\subsection {Topologies with Higgs associated production and \boldmath{$A/H\to\chi_1\chi_h,\, \chi_h\to\chi_1Z$}}
\label{ttZ}

In Sec.\,\ref{sec:framework}, we argued that the decay mode $A/H\to\chi_1\chi_h$, where $h=2,3$, can be one of the most promising heavy Higgs decay channels in the presence of light binos and Higgsinos. In Sec.\,\ref{Sec:neutralinos} we demonstrated that gluon fusion production of the heavy Higgs bosons followed by the above decay can be a promising topology for the HL-LHC, with very good prospects to probe sizable regions of parameter space in the $m_A-$tan\,$\beta$ plane. Here, we briefly discuss this decay topology with alternate heavy Higgs production mechanisms. 

{\bf{Large \boldmath{$\tan\beta$}}:} Heavy Higgs associated production with bottom quarks can have large cross sections ($\mathcal O(100~{\rm{fb}})$ at the boundaries of the $H\to\tau\tau$ experimental exclusion). The process $pp\to b\bar{b}(A/H)$, followed by $A/H\to\chi_1\chi_h,\, \chi_h\to\chi_1Z$, gives a $b\bar{b}Z+\met$ final state, with $\mathcal{O}(10)$ fb cross sections. This signature suffers from large backgrounds: multi-jet QCD background and $Z/W~+$ jets, if $Z$ decays hadronically; $t\bar{t}$ and $Z+$jets, if the $Z$ decays leptonically. It will be interesting to see if future LHC search strategies will be able to disentangle the signal from the very large background.

\begin{figure*}[t] 
 \vspace{0.cm}
  \center{\includegraphics[width=10cm]{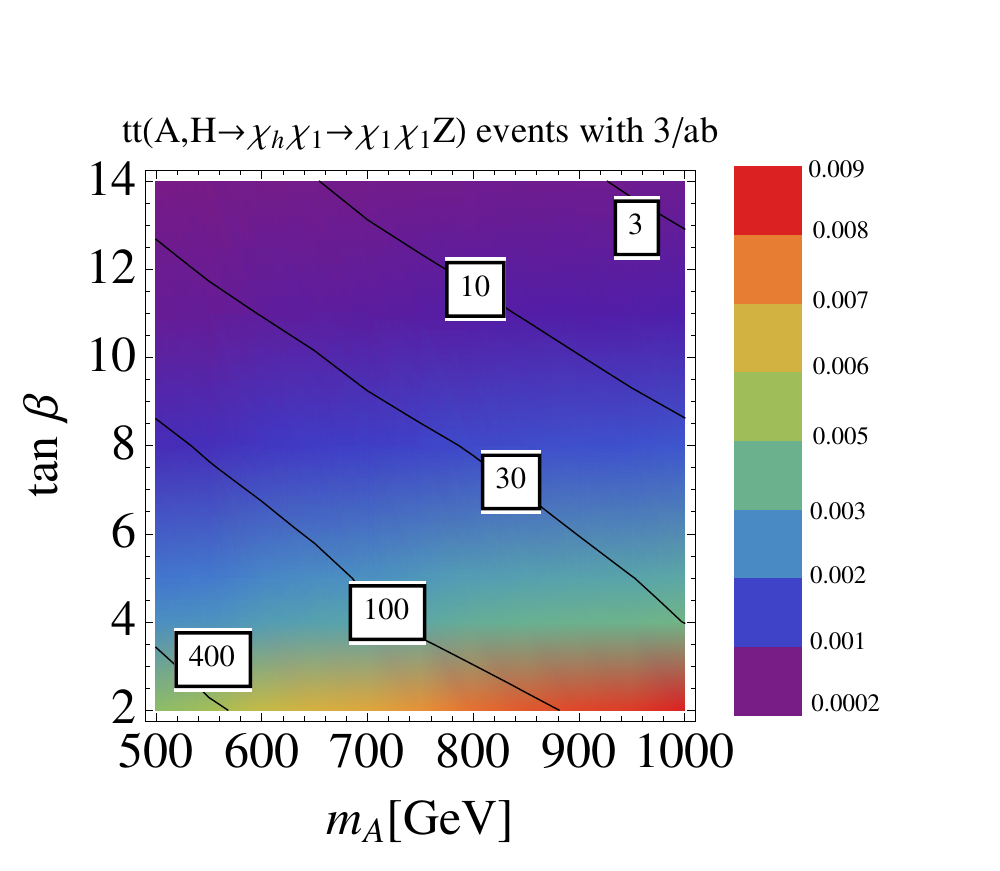}}
\caption{ Number of $pp\to t\bar{t}(A,H),A/H\to\chi_1\chi_h,\, \chi_h\to\chi_1Z$ events (summed over both $A$ and $H$) expected at the HL-LHC. Here we choose $\mu$ such that $m_{\chi_3}= m_A/2+25$ GeV, and $M_1$ such that $m_{\chi_1}= m_{\chi_3}-100$ GeV. The color coding shows the ratio of this production cross section to the cross section of mono-Z events from direct production of the electroweakino states, $pp\to\chi_i\chi_j\to Z+\met$.}
\label{Fig:ttZ}
\end{figure*} 

{\bf{Small \boldmath{$\tan\beta$}}:}  The $t\bar t H$ production cross section rises at low tan $\beta$, which is an unconstrained region of parameter space, and leads to the final state $t\bar{t}Z+\met$. Although this mode incurs a significant energy cost for the top pair production, it avoids the large QCD and $t\bar{t}$ backgrounds. Fig.\,\ref{Fig:ttZ} plots the number of $pp\to t\bar{t}(A,H),A,H\to\chi_1\chi_h,\, \chi_h\to\chi_1Z$ events (summed over both $A$ and $H$) expected at the HL-LHC. The color coding shows the ratio of this production cross section to the cross section of mono-Z events from direct production of the electroweakino states, $pp\to\chi_i\chi_j\to Z+\met$ (analogous to the color coding in Fig.\,\ref{Fig:gluonfusion}). Since the former production mode has to pay the hefty price of producing $t\bar{t}$ in addition to the heavy Higgs, we see that this cross section is always significantly smaller than the direct electroweakino production cross section. Nevertheless, hundreds of events are possible for this rich final state at the HL-LHC, hence this could be a challenging but viable additional search channel at small values of $\tan\beta$.

\subsection {Topologies with \boldmath{$A/H\to\chi_1\chi_h,\, \chi_h\to\chi_1h$ }}
\label{monoh}

If $m_{\chi_h}- m_{\chi_1}>m_h$, $\chi_h$ will also decay to $\chi_h\to\chi_1+h$ in addition to $\chi_h\to\chi_1+Z$. Hence all the final states discussed in the previous section can now involve the 125 GeV SM-like Higgs boson, $h$, in the final state instead of the $Z$-boson. 

\begin{figure*}[t] 
 \vspace{0.cm}
  \center{\includegraphics[width=10cm]{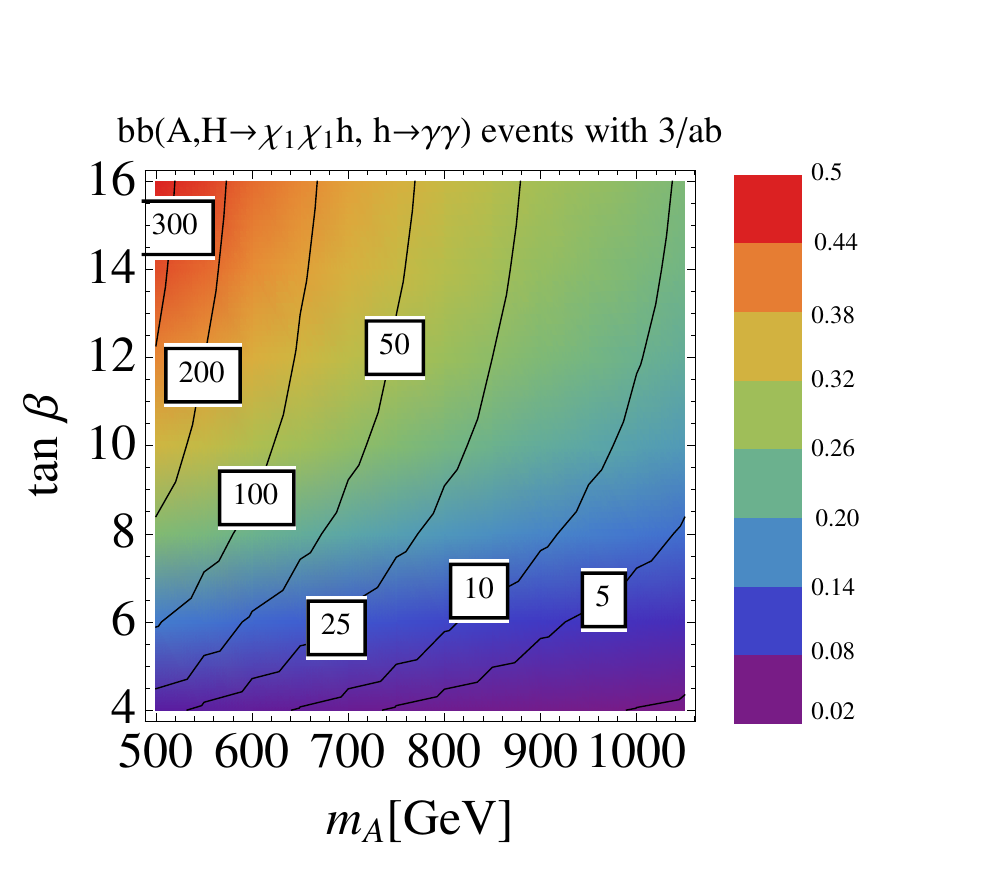}}
\caption{The number of $pp\rightarrow b \bar b(A,H), (A,H)\rightarrow \chi_1\chi_h$, $\chi_h\rightarrow \chi_1 h, h\to\gamma\gamma$ events (summed over $A$ and $H$) at the HL-LHC. For this plot, we choose the spectrum $\mu=m_A/2-10$ GeV, $M_1=\mu-130$ GeV, so that $A/H\to\chi_1\chi_h$ as well as $\chi_h\to\chi_1+h$ are kinematically open (for these parameters, we find $BR(\chi_h\to\chi_1h)\approx BR(\chi_h\to\chi_1Z)$, where $h$ represents a sum over $2,3$). The color coding shows the ratio of this production cross section (not including the BR$(h\to\gamma\gamma$)) to the production cross section of mono-h final states from direct electroweakino production, $pp\to\chi_i\chi_j\to h+\met$. }
\label{Fig:bbh}
\end{figure*}

For gluon fusion production of the heavy Higgs bosons, this leads to a mono-Higgs+$\met$ final state. The reach for such a decay topology has been discussed in \cite{Ellwanger:2017skc,Baum:2017gbj,Baum:2018zhf,Baum:2019uzg}, and was found to be weaker than the corresponding reach from the mono-Z signal for comparable branching ratios in the two channels. For b-associated production, several interesting signatures arise from the various Higgs decay channels such as $b\bar b, \gamma\gamma, WW, ZZ$. Of these, one of the cleanest is $h\to\gamma\gamma$, leading to a $2b+2\gamma+\met$ final state. In Fig.\,\ref{Fig:bbh}, we show the combined number of events from $pp\to b\bar{b}(A,H),\,(A,H)\to\chi_1\chi_h,\, \chi_h\to\chi_1h,\,h\to\gamma\gamma$ expected at the HL-LHC, with color coding showing the ratio of this production cross section (not including BR$(h\to\gamma\gamma$)) to the cross section of mono-h events from direct production of electroweakinos ($pp\to\chi_i\chi_j\to h+\met$). The production cross section for the topology in question is smaller than that for direct production. Nevertheless, $\mathcal O(100)$ events are possible at the HL-LHC. 

\subsection{Topologies involving \boldmath $A/H\to$ invisible} 
\label{invisible}

A potentially interesting signature arises when the heavy Higgs bosons decay invisibly. This occurs for $A/H\to\chi_1\chi_1$\footnote{For an analysis of heavy Higgs pair production with the Higgs decaying invisibly, see e.g. \cite{Arganda:2017wjh}.}, which, despite being suppressed by the bino-Higgsino mixing angle in our benchmark models, can have branching ratios of a few percent (see the first two panels of Fig.\,\ref{Fig:branchingratios}). This invisible decay channel leads to events with large $\met$, which can be tagged, for instance, with initial state radiation or b/t-associated production. Prospects for such signals were studied in \cite{Craig:2015jba}. For the bino-Higgsino benchmarks, even using the $\met$ handle, it is difficult to improve signal/background to better than percent level even at the HL-LHC for regions of parameter space currently not probed by other existing LHC searches. Thus, this is a somewhat challenging decay channel. This outlook might improve in other scenarios, for instance when winos are light.

\subsection{Topologies involving charged Higgs decays}
\label{chargedh}

As discussed in Sec.\,\ref{sec:framework}, the branching ratios of the charged Higgs boson into the $\chi^\pm\chi_i$ or $\tilde{\nu}\tilde{\tau}$  channels can be sizable (see right panel of Fig.\,\ref{Fig:branchingratios}). 

In Fig.\,\ref{Fig:chargedHiggs}, we show the number of events from the dominant production mode of charged Higgs, $pp\to tH^\pm X,\,H^\pm\to\chi^\pm\chi_1$\footnote{$X$ stands for either detector visible or invisible $b$ jets.}, at the HL-LHC. 
The color coding shows the branching ratio BR($H^+\to\chi^+\chi_1$). 
As discussed below Eq.\,(\ref{eq:Hchichi}), the coupling $H^+\chi^+\chi_1$ is largely independent of $\tan\beta$. Hence the branching ratio into this channel peaks around tan\,$\beta\sim7$ (as shown in the figure), where neither the top- nor bottom-type Yukawas are too large. On the other hand, the $tH^+b$ coupling, which controls the production cross section, is given by $\sqrt{2}(y_t P_R$ cot\,$\beta + y_b P_L$ tan\,$\beta$) and is large at small or large values of tan\,$\beta$. These two effects effectively balance out, so that the cross section $\sigma(pp\to tH^\pm X,\,H^\pm\to\chi^\pm\chi_1)$, and subsequently the expected number of events, is essentially independent of the value of tan\,$\beta$, as demonstrated by the contours. We find that $\mathcal{O}(1)$ fb cross sections are possible for $m_A\lesssim 700$ GeV, resulting in several thousand events at the HL-LHC.

\begin{figure*}[t] 
 \vspace{0.cm}
  \center{\includegraphics[width=10cm]{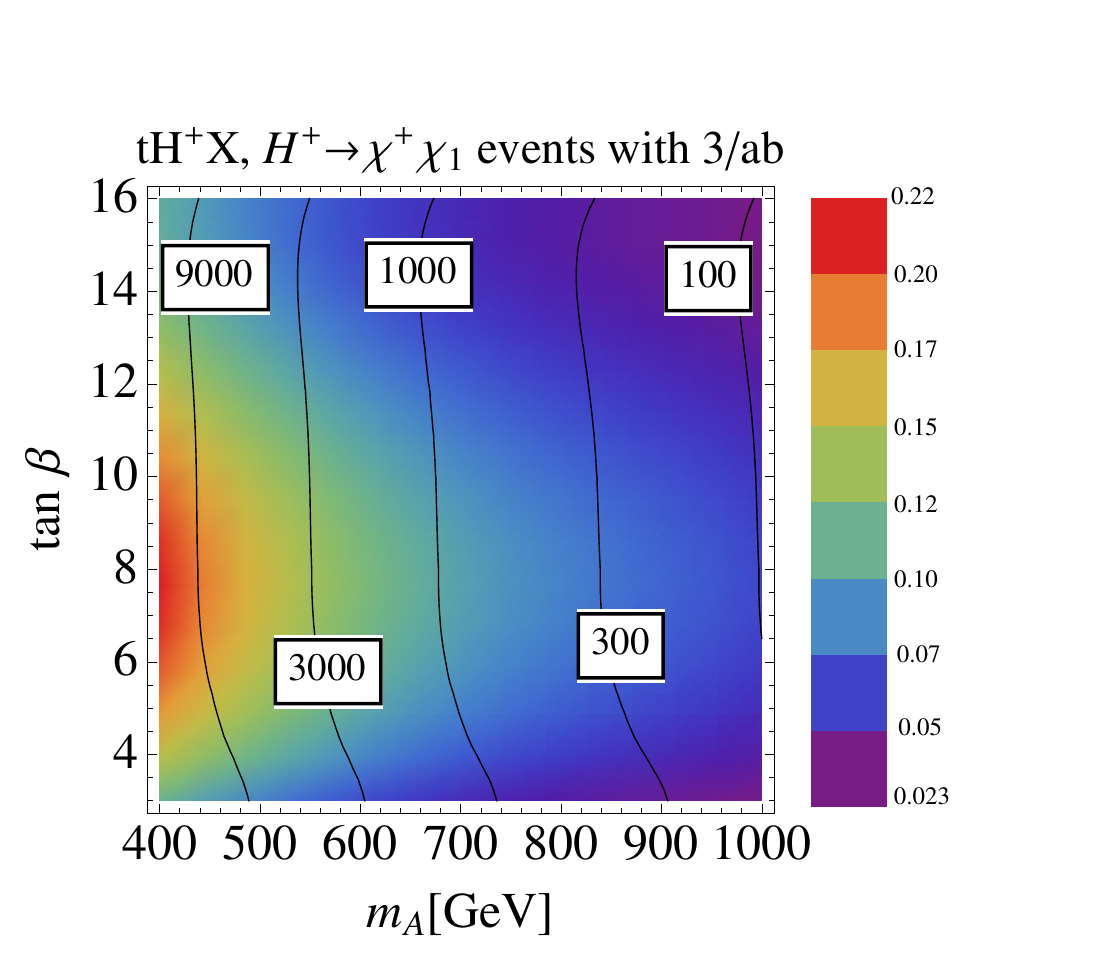}}
\caption{The number of events for $pp\to tH^\pm X,\,H^\pm\to\chi^\pm\chi_1$ at the HL-LHC. This plot uses the same mass spectrum as of Fig.\,\ref{Fig:Branchingratios} (left panel): in particular, $M_1=150$ GeV and $\mu=m_A-175$ GeV. The color coding denotes the branching ratio BR($H^+\to\chi^+\chi_1$).}
\label{Fig:chargedHiggs}
\end{figure*} 

The final state from this decay chain, $t b W+\met$, is challenging with $t\bar t$ providing the main background. However, as seen in the stau example in Section~\ref{Sec:staus}, heavy Higgs does provide additional kinematic handles that can beat down dominant multi-jet and $t\bar t$ background.
A different decay mode, such as $H^\pm\to\chi^\pm\chi_h$ ($h=2,3$), can evade this background thanks to the subsequent decay of $\chi_h$: $\chi_h\to Z/h\chi_1$, but BR($H^\pm\to\chi^\pm\chi_h$) is much smaller due to suppression of the coupling by the subleading bino-component of $\chi_h$ (see discussion below Eq.\,(\ref{eq:Hchichi})). The branching ratio can be enhanced in scenarios in which winos are not decoupled from the spectrum.

Similarly, the branching ratio into $\tilde{\nu}\tilde{\tau}$ can also be non-negligible. In the scenarios we discussed in Sec.\,\ref{Sec:staus}, the full decay chain arising from this decay mode is $pp\to tXH^\pm, H^\pm \to\tilde \tau\tilde\nu, \tilde\tau\to\tau+\met, \tilde\nu\to\nu+\met$, resulting in a $tb\tau+\met$ signature, with cross sections $\sim\mathcal{O}$(fb) at most. More interesting signals can emerge from a large mass splitting between the sneutrino and the stau, which can occur from a large splitting between the two stau soft masses and/or with a large trilinear $A_\tau$ and/or $\mu$ terms. This spectrum can lead to a sizable branching ratio for $\tilde\nu\to\tilde\tau W,\,\tilde\tau\to\tau+\met$ \cite{Carena:2012gp}, resulting in a rich $tb\tau\tau W+\met$ signature.

\section{Summary and discussion}
\label{Sec:summary}
In this paper, we have demonstrated that relatively generic SUSY spectra lead to sizable decay rates of heavy Higgs bosons into SUSY electroweak particles. These decay modes have not yet been explored by the LHC collaborations. The goal of the paper was twofold: (1) to propose new heavy Higgs search strategies for the LHC, and (2) to provide benchmark scenarios for the interpretation of future heavy Higgs searches. We have focused on models that are hidden to the standard searches at the LHC.
In the neutralino sector, we considered light binos and Higgsinos, which have small direct production cross sections, but decoupled the wino, which can be produced more copiously. Likewise, we considered light staus, which are notoriously difficult to probe directly at the LHC. 

For such hidden scenarios, in Section \ref{sec:framework} we systematically studied the possible interactions and branching ratios from heavy Higgs bosons, finding that heavy Higgs decays can be the dominant production mode of these particles at the LHC in some cases, and identified unsuppressed decay channels that are promising for searches. In particular, we identified two promising directions for new LHC searches:

(1) In Section \ref{Sec:neutralinos}, we studied search strategies for heavy Higgs decays into neutralinos via the process $pp\to A,H \to \chi_1\chi_h,\,\chi_h\to \chi_1 Z$ (h=2,3), which yields a $Z+\met$ signal. Making use of a modified clustered transverse mass of the final states, in conjunction with the large missing energy present in this process, we found that this decay channel can significantly extend the reach for heavy Higgs bosons in the intermediate tan\,$\beta\sim 2-8$ regime, a region of parameter space generally inaccessible from standard heavy Higgs searches (Fig.\,\ref{Fig:chimatbreach}).

(2) In Section \ref{Sec:staus}, we studied heavy Higgs decay to staus via b-associated production $pp\to b \bar b(A,H), A/H\to \tilde{\tau}\tilde{\tau},\,\tilde{\tau}\to\tau\chi_1$. We made use of the similarities to the $A/H\to\tau\tau$ searches, which provide the strongest limits for heavy Higgs bosons at large values of $\tan\beta$, and implemented additional kinematic cuts to make use of the heavy Higgs decay topology of the signal. We found that this channel can provide reach in the $(m_A-\tan\beta)$ plane comparable to the $\tau\tau$ search, suggesting that this channel can serve as a complementary channel in the discovery of heavy Higgs bosons. Even more importantly, such decays provide a novel search avenue for staus, which are difficult to probe directly even at HL-LHC. Heavy Higgs decays provide the dominant production mechanism for staus in large regions of parameter space (by over an order of magnitude in some cases, see Fig.\,\ref{Fig:stauCS}), offering possibilities for probing $m_{\tilde{\tau}}$ up to several hundred GeV (Fig.\,\ref{Fig:staulimits}).

In Section\,\ref{Sec:others}, we provided a brief overview of additional heavy Higgs decay channels that can be promising at the HL-LHC. 
It will be very interesting to see if improved search strategies will be able to extract such signatures in the future. A summary of the various channels that were discussed in this paper is presented in Table \ref{Tab:summary}. 

\begin{table}[t]
\begin{center}
\begin{tabular}{|l|l|l|}
\hline
\begin{tabular}[c]{@{}l@{}}Decay \\ Channel\end{tabular} & \begin{tabular}[c]{@{}l@{}}Production \\ Mode\end{tabular}                      & Comments                                                      \\ \hline\hline
\multicolumn{3}{|l|}{~~~~~Neutral Higgs bosons, A,H}                                                                                                                                                             \\ \hline
\multicolumn{3}{|l|}{decays to $\chi_{2,3} \chi_{2,3}, \chi_1\chi_1$ suppressed (however, see Sec.\,\ref{invisible}), $\chi^+\chi^-$ vanishes}                                                                                                                         \\ \hline
\multirow{3}{*}{$\chi_{2,3}\chi_1$}                               & gluon fusion                                                                    & $Z+\met$ at low/intermediate tan\,$\beta$ (detailed collider analysis in Sec.\,\ref{Sec:neutralinos})                 \\ \cline{2-3} 
                                                         & $b\bar bH$                                                                             & $b \bar bh$+$\met$ at large tan\,$\beta$ (Sec.\,\ref{monoh}) \\ \cline{2-3} 
                                                         & $t\bar tH$                                                                            & $t\bar tZ$ at small tan\,$\beta$ (Sec.\,\ref{ttZ})                                      \\ \hline
$\tilde{\tau}\tilde{\tau}$                                                & \begin{tabular}[c]{@{}l@{}}gluon fusion,\\ $b\bar bH$\end{tabular} & \begin{tabular}[c]{@{}l@{}}similar to $\tau\tau$ channel; best probed at intermediate/large tan\,$\beta$ \\ $b\bar b\tau\tau+\met$ (detailed analysis in Sec.\,\ref{Sec:staus})\end{tabular} \\ \hline\hline
\multicolumn{3}{|l|}{~~~~~Charged Higgs bosons, $H^{\pm}$}                                                                                                                                                                  \\ \hline
\multicolumn{3}{|l|}{decays to $\chi^\pm\chi_{2,3}$ suppressed}                                                                                                                                                                  \\ \hline
$\chi^\pm\chi_1$                                                 & $tH^\pm X$                                                                            & $tbW+\met$, no strong dependence on tan\,$\beta$ (Sec.\,\ref{chargedh})                                                      \\ \hline
$\tilde{\tau}\tilde{\nu}$                                           & $tH^\pm X$                                                                           & $tb\tau+\met$, large tan\,$\beta$                       (Sec.\,\ref{chargedh})                                 \\ \hline
\end{tabular}
\caption{Summary of heavy Higgs boson decay channels and HL-LHC prospects discussed in this work for scenarios with light binos-Higgsinos, or light staus. Here, ``suppressed" implies suppression of the vertex from the bino-Higgsino mixing angle.
}
\end{center}
\label{Tab:summary}
\end{table}

In conclusion, the LHC coverage of the electroweak sector of the MSSM has significant gaps where 
supersymmetric particles can be light and within reach of the LHC, but suffer from small production cross section and large SM background. In this paper, we have illustrated that making use of new production modes (heavy Higgs decays) and additional kinematic information available in the decay topology can significantly enhance the reach for such regions of parameter space, aiding the sensitivity to these superpartners as well as heavy Higgs bosons. As upcoming runs of the LHC gather data in the coming years, it is crucial to fully explore such possibilities in order to diversify and maximize the reach for heavy Higgs bosons, as well as supersymmetry, at the LHC. 

\section*{Acknowledgements}

We would like to thank Stefan Prestel and Maxim Perelstein for helpful discussions. SG is supported by the NSF CAREER grant PHY-1915852. BS is partially supported by the NSF CAREER grant PHY-1915852. The authors thank the Aspen Center for Physics, which is supported by National Science Foundation grant PHY-1607611.  This manuscript has been authored by Fermi Research Alliance, LLC under Contract No. DE-AC02-07CH11359 with the U.S. Department of Energy, Office of Science, Office of High Energy Physics. ZL is supported in part by the NSF under Grant No. PHY1620074 and by the Maryland Center for Fundamental Physics.

\bibliographystyle{utphys}
\bibliography{mssmhewkinobib}

\end{document}